\documentclass[12pt]{article}
 
\newlength{\abstractwidth}
\usepackage{amsmath, nccmath}
\usepackage{amsfonts}
\usepackage{amssymb}
\usepackage{latexsym}

\usepackage{color}
\usepackage{graphicx}
\usepackage{tikz}
\usepackage{dsfont}

\usetikzlibrary{arrows,shapes,positioning}
\usetikzlibrary{decorations.markings}
\usepackage[rightcaption]{sidecap}
\tikzstyle arrowstyle=[scale=1]
\tikzstyle directed=[postaction={decorate,decoration={markings,
    mark=at position .65 with {\arrow[arrowstyle]{stealth}}}}]
\tikzstyle reverse directed=[postaction={decorate,decoration={markings,
    mark=at position .65 with {\arrowreversed[arrowstyle]{stealth};}}}]
\usetikzlibrary{positioning}

\usepackage{hyperref}
\definecolor{darkred}{rgb}{0.8,0.1,0.1}
\hypersetup{colorlinks=true, linkcolor=darkred, citecolor=blue, linktoc=page}

\numberwithin{equation}{section}

\renewcommand{\thefootnote}{\fnsymbol{footnote}}
\renewcommand{\thanks}[1]{\footnote{#1}}
\newcommand{\starttext}{
\setcounter{footnote}{0}
\setcounter{section}{0}
\renewcommand{\thefootnote}{\arabic{footnote}}}
\newcommand{\bea}{\begin{eqnarray}}
\newcommand{\eea}{\end{eqnarray}}
\newcommand{\be}{\begin{eqnarray}}
\newcommand{\ee}{\end{eqnarray}}
\newcommand{\bma}{\begin{matrix}}
\newcommand{\ema}{\cr\end{matrix}}

\newtheorem{thm}{Theorem}[section]
\newtheorem{lem}[thm]{Lemma}

\def\cA{{\cal A}}
\def\cB{{\cal B}}

\def\cD{{\cal D}}

\def\cG{{\cal G}}
\def\cH{{\cal H}}

\def\cM{{\cal M}}

\def\cR{{\cal R}}

\def\cT{{\cal T}}
\def\cU{{\cal U}}

\def\cZ{{\cal Z}}

\def\mA{\mathfrak{A}}
\def\mB{\mathfrak{B}}

\def\mJ{\mathfrak{J}}

\def\mM{\mathfrak{M}}

\def\ZZ{{\mathbb Z}}
\def\RR{{\mathbb R}}

\def\CC{{\mathbb C}}

\def\Im{{\rm Im \,}}

\def\p{\partial}

\def\a{\alpha}
\def\b{\beta}
\def\g{\gamma}

\def\f{\varphi}

\def\ep{\varepsilon}
\def\om{\omega}

\def\pbz{\p _{\bar z}}

\def\beq{\begin{equation}}
\def\eeq{\end{equation}}

\def\pbx{\p _{\bar x}}
\def\pby{\p _{\bar y}}

\def\pbz{\p _{\bar z}}
\def\GA{\cG}
\def\kap{\kappa}
\def\oom{\overline{\om}}

\def\barJ{{\bar J}}

\def\no{\nonumber}
\def\sm{\smallskip}

\allowdisplaybreaks

\begin{document}
\starttext
\setcounter{footnote}{0}

\begin{flushright}
2020 December 19\\
revised 2021 September 12 \\
UUITP-36/20 \\
\end{flushright}

\bigskip

\begin{center}

{\Large \bf Identities among higher genus modular graph tensors}

\vskip 0.4in

{  \bf Eric D'Hoker$^{(a)}$, and Oliver Schlotterer$^{(b)}$}

\vskip 0.1in

 ${}^{(a)}$ {\sl Mani L. Bhaumik Institute for Theoretical Physics}\\
 { \sl Department of Physics and Astronomy }\\
{\sl University of California, Los Angeles, CA 90095, USA}\\

\vskip 0.1in

 ${}^{(b)}$ { \sl Department of Physics and Astronomy,} \\ {\sl Uppsala University, 75108 Uppsala, Sweden}
 
 \vskip 0.1in
 
{\tt \small dhoker@physics.ucla.edu},  {\tt \small oliver.schlotterer@physics.uu.se}

\begin{abstract}
Higher genus modular graph tensors map Feynman graphs to functions on the Torelli space of genus-$h$ compact Riemann surfaces  which transform as tensors under the modular group $Sp(2h , \ZZ)$, thereby generalizing a construction of Kawazumi. An infinite family of algebraic identities between one-loop and tree-level modular graph tensors are proven for arbitrary genus and arbitrary tensorial rank. We also derive a family of identities that apply to modular graph tensors of higher loop order.
\end{abstract}

\end{center}

\newpage

\setcounter{tocdepth}{2} 

\newpage

\baselineskip=16pt
\setcounter{equation}{0}
\setcounter{footnote}{0}

\section{Introduction}
\label{sec:1}
\setcounter{equation}{0}

Modular graph functions map Feynman graphs for a massless scalar field on a Riemann surface to a modular function or, more generally, to a modular form. Modular graph functions and forms arise naturally in the low energy expansion of closed string amplitudes as the integrands on moduli space of the coefficients of the low energy effective interactions. 

\sm

The study of genus-one modular graph functions goes back to \cite{Green:1999pv,Green:2008uj} where individual contributions at low order in the expansion were considered. A more general analysis of genus-one modular graph functions was initiated in  \cite{D'Hoker:2015foa,Zerbini:2015rss,DHoker:2015wxz} where they were found to satisfy identities that extend the well-known relations between multiple zeta-values to modular functions. Algorithms for the systematic  construction of all algebraic and differential identities between modular graph functions and forms were developed and applied in \cite{DHoker:2016mwo,Basu:2016kli,DHoker:2016quv, Gerken:2018zcy}, expressed in terms of generating functions for iterated integrals of holomorphic Eisenstein series in \cite{Gerken:2019cxz,Gerken:2020yii}, and implemented in a convenient {\sc Mathematica} package in \cite{Gerken:2020aju}. Further mathematical developments in the study of modular graph forms may be found in \cite{Brown:2017qwo, Brown2}.

\sm

The study of genus-two modular graph functions \cite{DHoker:2017pvk, DHoker:2018mys} is motivated by the analysis of the low energy expansion of the genus-two superstring amplitudes with four external massless states in \cite{D'Hoker:2002gw,D'Hoker:2005jc,Berkovits:2005df}.
The string integrand for the coefficient  of the $D^4 \cR^4$  effective interaction is a constant on moduli space in \cite{DHoker:2005jhf}, while for $D^6 \cR^4$ it is proportional to the genus-two Kawazumi-Zhang invariant \cite{D'Hoker:2013eea} which had been introduced a few years earlier in the mathematics literature \cite{Kawazumi,Zhang}. Using Siegel's formula for the volume of moduli space and an inhomogeneous  Laplace eigenvalue equation for the Kawazumi-Zhang invariant derived in \cite{DHoker:2014oxd} (see also \cite{Kawazumi}), the integrations over moduli space could be performed analytically. The resulting coefficients were found to match the predictions of space-time supersymmetry \cite{Green:1998by} and S-duality \cite{Green:1997tv,Green:1999pu,Green:2005ba}
 in Type IIB string theory \cite{DHoker:2005jhf} and \cite{DHoker:2014oxd}. A theta-lift for the Kawazumi-Zhang invariant was constructed in \cite{Pioline:2015qha}.

\sm

The recent construction of genus-two five-point superstring amplitudes \cite{DHoker:2020prr,DHoker:2020tcq,DMS3}  introduces, via their low energy expansion, additional classes of modular graph functions other than those already known from four-point amplitudes in \cite{DHoker:2017pvk, DHoker:2018mys}. The modular graph functions in the $D^8 \cR^4$ and $D^6 \cR^5$ effective interactions were found to obey an intriguing algebraic identity that mixes genus-two contributions from Feynman graphs of different loop orders  \cite{DHoker:2020tcq}. This identity was proven with the help of  interchange Lemma D.1 in \cite{DHoker:2020tcq}, which prescribes a certain interchange rule for derivatives on higher genus Arakelov Green functions similar to the interchange formula resulting from translation invariance at genus one. This Lemma will be exploited here as well and further generalized. The algebraic  identity of \cite{DHoker:2020tcq}  is closely related to a differential relation obtained earlier in \cite{Basu:2018bde}.

\sm

In this work, we shall extend the notion of genus-$h$ modular graph functions to modular graph tensors transforming under the modular group $Sp(2h,\ZZ)$. Genus-one  modular graph tensors reduce to modular graph forms, while for higher genus they generalize a construction of Kawazumi in \cite{Kawa1, Kawa2}. Based on concatenations of Arakelov Green functions, we recursively construct tensorial functions on one or two copies 
of the compact Riemann surface.
The associated generalizations of the interchange lemma \cite{DHoker:2020tcq} are used to
derive infinite classes of identities among modular graph tensors that arise from integrating the 
concatenated Green functions over the surface. More specifically, modular tensors corresponding to one-loop graphs with an arbitrary number of edges are related to modular tensors corresponding to linear tree-level graphs.

\subsection*{Acknowledgments}

We gratefully acknowledge our collaboration with Michael Green and Boris Pioline on earlier projects which inspired this work. Moreover, we are indebted to one of the referees for valuable suggestions on the manuscript. The research of ED is supported in part by NSF grant PHY-19-14412. The research of OS  is supported by the European Research Council under ERC-STG-804286 UNISCAMP.

\newpage

\section{The interchange lemmas}
\label{sec:2}
\setcounter{equation}{0}

In this section, we begin by fixing notations and reviewing the ingredients needed to construct modular graph tensors for arbitrary genus $h$, namely holomorphic Abelian differentials and the Arakelov Green function (for reviews see \cite{fay73,D'Hoker:1988ta}). We shall then summarize the  interchange lemma proven in \cite{DHoker:2020tcq}, and prove its generalization to higher rank tensors.

\subsection{Holomorphic one-forms}

We consider a compact Riemann surface $\Sigma$ of genus $h>0$ and choose a basis of cycles $\mA_I$ and $\mB_I$  in $H_1(\Sigma , \ZZ)$ for which the intersection pairing $\mJ$ takes the canonical form $\mJ(\mA_I, \mA_J)  = \mJ(\mB_I, \mB_J)  =  0$ and  $\mJ(\mA_I, \mB_J)  =  \delta _{IJ}$ for $I,J =1,\cdots, h$. A canonical basis of holomorphic Abelian differentials $\om_I$ for $H^{(1,0)} (\Sigma)$ may be normalized  on $\mA$-cycles, and we have, 
\bea
\label{omdef}
\oint _{\mA_I} \om_J = \delta _{IJ} 
\hskip 1in 
\oint _{\mB_I} \om_J = \Omega _{IJ} 
\hskip 1 in Y_{IJ} = \Im \Omega _{IJ}
\eea
The period matrix $\Omega$ is symmetric and has positive definite imaginary part $Y$. The action of a modular transformation  $\mM\in Sp(2h,\ZZ)$ on the periods of $\mA_I$ and $\mB_I$ preserves the canonical intersection form, namely $\mM^t \, \mJ \, \mM= \mJ$ with,
\bea
\label{mod1}
 \left ( \begin{matrix} \tilde \mB \cr \tilde \mA \cr \end{matrix} \right ) 
 = \mM \left ( \begin{matrix} \mB \cr \mA \cr \end{matrix} \right )
 \hskip 0.7in 
\mJ = \left ( \begin{matrix}0 & -I_h \cr I_h & 0 \cr \end{matrix} \right )
 \hskip 0.7in
\mM = \left ( \begin{matrix} A & B \cr C & D \cr \end{matrix} \right )
\eea
where $A,B,C,D$ are $h \times h$ matrices with integer entries. The action of the modular transformation $\mM$ on the row matrix  $\om$ of $h$ holomorphic differentials, on the period matrix $\Omega$, and on the  imaginary part of the period matrix $Y$ are given by,
\bea
\label{mod2}
\tilde \om & = & \om (C \Omega+D)^{-1}
\no \\
\tilde \Omega & = & (A \Omega +B) (C \Omega +D)^{-1} 
\no \\
\tilde Y & = & (\Omega C^t+D^t)^{-1} Y (C \bar \Omega +D)^{-1}
\eea
The Jacobian variety $J(\Sigma)= \CC^h/(\ZZ^h + \Omega \ZZ^h)$ supports a canonical translation-invariant K\"ahler form whose pull-back from $J(\Sigma)$ to $\Sigma$ under the Abel-Jacobi map induces a canonical conformal invariant K\"ahler form on $\Sigma$ given by (with $Y^{IJ}$ denoting the entries of $(\Im \Omega)^{-1}$),\footnote{Throughout, we  use the Einstein convention to contract pairs of repeated upper and lower indices; upper indices are lowered with the help of the matrix $Y$ with entries $Y_{IJ}$;  lower indices are raised with the inverse matrix $Y^{-1}$ with entries $Y^{IJ}$; and we shall reserve lower indices on holomorphic forms $\om_I$ and upper indices for their complex conjugates $\oom^I$. }
\bea
\label{kappa}
\kap   = { i \over 2h}  \omega_I \, \oom^I 
\hskip 1in 
\oom ^I = Y^{IJ} \oom _J
\hskip 1in 
\int _\Sigma \kappa =1
\eea

\subsection{Modular tensors}

The moduli space of compact Riemann surfaces equipped with a canonical homology  basis  $(\Sigma, \mA, \mB)$ is the Torelli space ${\rm Tor}_h$, which is a subspace of the Siegel upper half space $\cH_h$ of rank $h$. The moduli space of compact Riemann surfaces is isomorphic to the quotient $\cM_h = {\rm Tor}_h/ Sp(2h,\ZZ)$. The Siegel upper half space $\cH_h$ is a K\"ahler coset manifold given by $Sp(2h, \RR)/U(h)$. For $h=2$, and for $h >2$ away from the locus of hyper-elliptic Riemann surfaces, the K\"ahler structure of $\cH_h$ induces a K\"ahler structure on ${\rm Tor}_h$ thereby decomposing the tangent space of ${\rm Tor}_h$  into a direct sum of holomorphic and anti-holomorphic subspaces. For $h>2$, along the hyper-elliptic locus,  the tangent map of the inclusion of ${\rm Tor}_h$ into $\cH_h$ vanishes on the $(-1)$-eigenspace of the hyper-elliptic involution, so that the K\"ahler structure of $\cH_h$ does not extend to a non-degenerate K\"ahler structure on ${\rm Tor}_h$. This fact implies that all the tensors to be considered in the present paper degenerate in the normal direction of the hyper-elliptic locus.\footnote{We are very grateful to one of the referees for pointing out this behavior.} In particular, a vector transforming under the defining representation of $Sp(2h,\RR)$ may be decomposed into a vector $V_I$ with a holomorphic index $I$ and a vector $V_{\bar J}$ with an anti-holomorphic index $\bar J$. This decomposition induces a corresponding decomposition under the transformation of the modular group $Sp(2h,\ZZ) \subset Sp(2h, \RR)$.

\sm

The prototype of such a decomposition is given by the $2h$-component vector  $(\om_I, \oom_\barJ)$ of holomorphic 1-forms and their complex conjugates, and its transformation rule of  (\ref{mod2}) under $Sp(2h,\ZZ)$ may be written in complex tensor notation as follows,
\bea
\tilde \om_I (\Omega) = \om _{I'}(\Omega) \, R^{I'}{}_I \hskip 1in 
R= R(\Omega) = (C \Omega + D)^{-1}
\eea
Higher rank tensors may be obtained by repeated tensor products of the defining representation, and decomposed analogously. The transformation law for the rank-two tensor $Y_{I {\bar J}}$ under $Sp(2h,\ZZ)$ was already given in (\ref{mod2}).\footnote{To emphasize the K\"ahler structure of the tensors on ${\rm Tor}_h$, we shall distinguish here between indices $I$ and $\bar J$ corresponding to holomorphic tangent space directions and their complex conjugates.} The transformation for an arbitrary tensor-valued function $\cT(\Omega)$ is given as follows, 
\bea
\tilde \cT_{I_1, \cdots, I_n; \barJ_1,  \cdots , \barJ_{\bar n}} (\tilde \Omega) 
=  \cT_{I_1',  \cdots, I_n';\barJ_1', \cdots , \barJ_{\bar n}'} (\Omega) \, 
R^{I_1'}{}_{I_1}  \cdots R^{I_n'}{}_{I_n} \, 
\bar R^{\barJ_1'}{}_{\barJ_1}  \cdots  \bar R^{\barJ_{\bar n}'}{}_{\barJ_{\bar n}}
\eea
Equivalently, we may raise the anti-holomorphic indices with the help of $Y^{-1}$, 
\bea
\cT_{~I_1, \cdots, I_n}^{ J_1,  \cdots , J_{\bar n}} (\Omega) 
= \cT_{I_1, \cdots, I_n; \barJ_1,  \cdots , \barJ_{\bar n}} (\Omega) \, Y^{J_1 \barJ_1} \cdots Y^{J_{\bar n} \barJ_{\bar n}}
\eea
whose transformation law is now by the holomorphic matrices $R(\Omega)$ and $R(\Omega)^{-1}$ without the need to invoke complex conjugates $\bar R(\Omega)$,
\bea
\label{Trans}
\tilde \cT_{~I_1, \cdots, I_n}^{ J_1,  \cdots , J_{\bar n}} (\tilde \Omega)
= \cT_{~I_1', \cdots, I_n'}^{ J_1',  \cdots , J_{\bar n}'} (\Omega) \, 
R^{I_1'}{}_{I_1}  \cdots R^{I_n'}{}_{I_n} \, 
(R^{-1}) ^{J_1}{}_{J_1'}  \cdots (R^{-1}) ^{J_{\bar n}}{}_{J_{\bar n}'} 
\eea
In the sequel, we shall construct modular graph tensors obtained as tensor-valued functions on ${\rm Tor}_h$ associated with tree-level and one-loop Feynman graphs, and the key ideas of our construction are applicable to graphs with an arbitrary number of loops.
Our formulation generalizes a construction of modular tensors initiated by Kawazumi in \cite{Kawa1} and \cite{Kawa2}.

\subsection{The Arakelov Green function}

The Arakelov Green function $\cG$ is a symmetric real-valued function on $\Sigma \times \Sigma$ defined as the unique inverse to the Laplace operator on the space of functions orthogonal to constants with respect to the volume form $\kappa$ and,   in local complex coordinates $x,y$, is defined by,
\bea
\label{Ara1}
\p_x \pbx \cG(x,y) = - \pi \delta (x,y) + \pi \kappa _{x \bar x} (x)
\hskip 1in
\int_{\Sigma_x} \kappa (x) \cG(x,y)=0
\eea
The $\delta$-function is normalized by ${ i \over 2} \int_{\Sigma_x} dx \wedge d \bar x \, \delta (x,y)=1$,
the subscript of $\Sigma_x$ specifies the integration variable when appropriate\footnote{We shall drop
the explicit reference to the integration variables and simply write $\int_\Sigma$ when all the
points on the surface appearing in the subsequent expression are integrated over.}, and $\kappa$ is given by $\kap    = { i \over 2} \kappa_{x \bar x} dx \wedge d\bar x$. An explicit form for $\cG$ may be  constructed in terms of the prime form and Abelian integrals \cite{DHoker:2017pvk}, but will not be needed in this paper. A formula that will be useful here is as follows,
\bea
\label{Ara2}
\p_x \pby \cG(x,y) = \pi \delta (x,y) - \pi \om_I(x) \, \oom^I(y)
\eea
Using the Riemann relation,
\bea
{ i \over 2} \int _\Sigma \om _I \, \oom ^J = \delta _I^J
\eea
one readily verifies that the integral of the above relation against an arbitrary holomorphic form $\om_K(y)$ vanishes, as does the integral against an arbitrary anti-holomorphic form $\oom^K(x)$.

\subsection{The basic interchange lemma}

Translation invariance on a compact Riemann surface of genus one, namely a torus,  guarantees that the holomorphic differential $\om_I$ is a constant and that the Arakelov Green function depends only on the difference of the points, so that $\cG(z,w) |_{h=1} = g(z-w)$ and $\p_z \cG(z,w) |_{h=1} = \p_z g(z-w) = - \p_w g(z-w) = - \p_w \cG(z,w) |_{h=1}$. These properties may be used to ``move derivatives around a Feynman graph" and are responsible for the momentum conservation identities relating modular graph forms of genus one \cite{DHoker:2016mwo,Basu:2016kli,DHoker:2016quv}. 

\sm

The absence of translation invariance on a Riemann surface $\Sigma$ of genus $h >1$ had prevented the derivation of corresponding identities until the basic interchange Lemma D.1 in  \cite{DHoker:2020tcq} was proven to provide a viable substitute. Its statement is as follows.
{\lem
\label{lemma1}
On a compact Riemann surface of  arbitrary genus $h \geq 1$, the following relation between derivatives of the Arakelov Green function $\cG$ and the Abelian differentials $\om_I$ holds,
\bea
\label{lemma1a}
\p_x W_I(x,y) = - \p_y W_I (y,x)
\eea
where the  tensor $W_I$ is given as follows, 
 \bea
 \label{lemma1b}
 W_I(x,y) & = & \cG(x,y) \om _I(y) - \Phi _I ^J (x) \om_J(y)
 \no \\
 \Phi_I^J (z) & = & {i \over 2} \int_{\Sigma_x} \GA(z,x) \, \om_I(x) \oom^J(x)
 \eea
Here, $W_I(x,y)$ is a $(0,0)$ form in $x$ and a $(1,0)$ form in $y$; $\Phi_I^J(x)$ is a $(0,0)$ form in $x$, 
traceless in $I,J$ by (\ref{Ara1}), and the matrix $\Phi$ is Hermitian as it satisfies $\overline{\Phi_{I \barJ} (x) } = \Phi _{J \bar I} (x)$.}

\sm

The tensor $\Phi_I^J(x)$ was introduced by Kawazumi in \cite{Kawa1,Kawa2} from a slightly different perspective.
The proof of the Lemma is given in Appendix D of \cite{DHoker:2020tcq}. For genus $h=1$ we have $\Phi_1^1(x)=0$ and $W_1(x,y)= \cG(x,y) = g(x-y)$ so that the equation for $W_I(x,y)$ becomes equivalent to the equation obtained by using translation invariance for genus one. For genus two, the lemma \ref{lemma1} is at the root of the proof of the identity between weight-two modular graph functions discovered in \cite{DHoker:2020tcq}. For arbitrary genus $h$, the lemma will serve as the starting point for all the identities we shall derive in this paper.

 \subsection{Concatenation of Arakelov Green functions}

Higher weight modular graph functions involve integrals of products of Arakelov Green functions over several copies of $\Sigma$. For genus one, these integrations are effected in terms of the unique canonical volume form ${ i \over 2} dz \wedge d\bar z$ on $\Sigma$ obtained by taking the wedge product of the canonical holomorphic $(1,0)$ form $dz$ and its complex conjugate. For higher genus $h \geq 2$, the space of holomorphic $(1,0)$ forms has dimension $h$ greater than one, so that several possible volume forms of the following type, 
\bea
\label{mu}
\mu_I ^J  = { i \over 2 } \om_I  \, \oom ^J  \hskip 1in \int _\Sigma \mu_I^J = \delta _I^J
\eea
for $ I,J=1, \cdots, h$, are available. While the canonical K\"ahler form $\kappa$, introduced in (\ref{kappa}) and related to $\mu$ by the trace $\mu _I^I = h \, \kappa$, is a natural modular invariant volume form on a single copy of $\Sigma$, the explicit expressions for higher-genus string amplitudes\footnote{More specifically, integration measures beyond several copies of $\kappa$ arise in genus-two string amplitudes with four external states \cite{D'Hoker:2005jc,Berkovits:2005df} and with five external states \cite{DHoker:2020prr,DHoker:2020tcq,DMS3} and the low energy limit of the genus-three four-point amplitude \cite{Gomez:2013sla}.} show that the measure on several copies of $\Sigma$ is not necessarily given by several copies of $\kappa$. Instead, the measures in higher-genus string amplitudes involve a more interrelated tensorial structure, which can be accounted for in terms of the tensorial form $\mu$ defined in (\ref{mu}).

\subsubsection{Concatenation of two Arakelov Green functions}

Following the hints given by the structure of the string amplitudes, we shall define a concatenation of two Green functions in terms of tensorial volume forms. Actually, there are two different perspectives of interest. The first is given by the straightforward concatenation of two Arakelov Green functions with the volume form $\mu_I^J$, while the second is in terms of the combination $W_I(x,y)$ used in Lemma \ref{lemma1}, 
\bea
\label{VW}
V_I^J (x,y) & = & \int _{\Sigma_z} \cG(x,z) \, \mu_I^J(z) \, \cG(z,y)
\no \\
W_{IL} ^J (x,y) & = & { i \over 2} \int_{\Sigma_z} W_I(x,z) \, \oom^J(z) \, W_L(z,y)
\eea
The tensor $V_I^J(x,y)$ generalizes the Arakelov Green function $\cG(x,y)$ as both are $(0,0)$ forms in both $x$ and $y$, while the tensor $W_{IL}^J(x,y)$ generalizes $W_I(x,y)$ as both are $(0,0)$ forms in $x$ and $(1,0)$ forms in $y$. 
While $V_I^J$ is a natural object to introduce, it is the combination $W_{IL}^J(x,y)$ that obeys the simplest generalization of Lemma \ref{lemma1}, and satisfies,
\bea
\label{W2}
\p_x W_{IL}^J(x,y) = - \p_y W^J_{LI} (y,x)
\eea
The proof of (\ref{W2}) follows immediately from the integral representation of $W_{IL}^J(x,y)$ in (\ref{VW}), the use of  lemma \ref{lemma1} applied to $\p_x W_I(x,z)$, integration by parts in $z$ and then a second use of the lemma \ref{lemma1} on $\p_z W_L(z,y)$. 

\sm

The relation between the two expressions may be obtained by substituting the expression for $W_I$ from (\ref{lemma1b}), and we find, 
\bea
\label{otherW2}
W_{IL}^J(x,y) = \Big ( V_I^J(x,y) - \Phi_I^K(x) \Phi_K^J(y) \Big ) \om_L(y) 
- \Big ( \Phi _{IL}^{JM} (x)  - \Phi_I^K(x) \cA_{KL}^{JM} \Big ) \om_M(y)
\eea
where the new ingredients are given as follows,
\bea
\Phi _{IL}^{JM} (x) & = & \int _{\Sigma_z} \cG(x,z) \, \mu_I^J(z) \, \Phi_L^M (z)
= \int _{\Sigma ^2_{z,w}}  \cG(x,z) \, \mu_I^J(z) \, \cG(z,w) \, \mu^M_L (w)
\no \\
\cA _{KL} ^{JM} & = & \int _{\Sigma_z} \mu_K^J (z) \Phi _L^M(z)
= \int _{\Sigma ^2_{x,y}} \mu_K^J (x) \, \cG(x,y) \, \mu_L^M (y)
\eea
The tensor $\Phi _{IL}^{JM} (x)$ generalizes $\Phi_I^J(x)$ by concatenation. The tensor $\cA_{KL}^{JM}$ was introduced by Kawazumi in \cite{Kawa1,Kawa2} and has no dependence on points on the surface $\Sigma$. It transforms as a tensor under $Sp(2h,\ZZ)$, according to (\ref{Trans}) with $n=\bar n=2$. Its symmetry and trace properties are as follows, 
\bea
\cA _{KL} ^{JM} = \cA _{LK} ^{MJ} 
\hskip 0.5in 
\cA _{KL} ^{JL} = \cA _{JL} ^{JM} = 0
\hskip 0.5in
\cA _{KJ} ^{JK} = \f
\eea
where $\f$ is the Kawazumi-Zhang invariant for arbitrary genus.

\sm

For later use, we record the following relations, 
\bea
\p_x \pby V^J_I(x,y) & = & 
\p_x \Phi _I^\a(x) \, \pby \Phi _\a^J (y) + \pi W_I(y,x) \, \oom^J(y)
\no \\ &&
- \pi \om_I(x) \, \oom^\a(y) \, \Phi^J_\a(x) + \pi \om_\b(x) \, \cA_{I\a} ^{\b J} \, \oom^\a(y)
\no \\
\p_x \pby W^J_{IL}(x,y) & = & - 2 \pi i \,  W_I(y,x) \, \mu_L^J(y)
+ 2 \pi i \, \om_I(x) \, \Phi^J_\a(x) \, \mu_L^\a (y) 
\no \\ &&
 - 2 \pi i \, \om_\b(x) \, \cA_{I\a} ^{\b J} \,  \mu_L^\a (y) 
\eea
For the sake of extra clarity, we shall frequently denote pairs of contracted indices by lower case Greek letters.

\subsubsection{Concatenation of an arbitrary number of Arakelov Green functions}

The generalization to the concatenation of an arbitrary number of Arakelov Green functions, following the pattern given above for the case of two Green functions, is straightforward. The corresponding tensor functions are defined recursively as follows for $n \geq 1$,
\bea
\label{VWn}
V_{\, I_1 \cdots \, I_n \, I}^{J_1 \cdots J_n \, J}  (x,y) & = & 
\int _{\Sigma_z} V_{\,I_1 \cdots \, I_n }^{J_1 \cdots J_n }  (x,z) \, \mu_I^J(z) \, \cG(z,y)
\no \\
W_{I_1 \cdots I_n \, I \, L} ^{J_1 \cdots J_n \, J}  (x,y) & = & 
{ i \over 2} \int _{\Sigma_z} W_{I_1 \cdots I_n \, I} ^{J_1 \cdots J_n }  (x,z)  \, \oom^J(z) \, W_L(z,y)
\no \\
\Phi _{I \, I_1 \cdots I_n } ^{J \, J_1 \cdots J_n} (x) & = &
\int _{\Sigma_z} \cG(x,z) \, \mu_I^J(z) \, \Phi _{I_1 \cdots I_n } ^{J_1 \cdots J_n} (z) 
\no \\
\cA_{I \, I_1 \cdots I_n} ^{J \, J_1 \cdots J_n} & = & \int _{\Sigma_z} \mu_I^J(z) \, \Phi _{I_1 \cdots I_n } ^{J_1 \cdots J_n} (z) 
\eea
The tensors  have the following symmetry properties,
\bea
V_{\, I_n \cdots  \, I_1 }^{J_n \cdots J_1 }  (x,y) & = & V_{\, I_1 \cdots \, I_n }^{J_1 \cdots J_n }  (y,x)
\no \\
\cA_{I_n \cdots I_1} ^{J_n \cdots J_1} & = & \cA_{I_1 \cdots I_n} ^{J_1 \cdots J_n} 
\label{reflec}
\eea
and trace properties,
\bea
\Phi_{I_1 \cdots I_n \, K}^{J_1 \cdots J_n \, K}(x)= \cA_{I_1 \cdots I_n \, K} ^{J_1 \cdots J_n \, K}=0
\eea
They are tensors under $Sp(2h,\ZZ)$ with transformation properties induced by those of $\om_I$ and $\oom^J$ and given by (\ref{Trans}) with $\bar n = n$. The other tensors have similar transformation properties. We shall refer to the objects $ \cA^{J_1 \cdots J_n}_{I_1 \cdots I_n}$ as {\sl modular graph tensors}.

\sm

We note that the recursion relations for $V$ and $W$ may also be written in opposite order, 
\bea
\label{VWnn}
V_{ \, I_1 \cdots \, I_n \, I}^{J_1 \cdots J_n \, J}  (x,y) & = & 
\int _{\Sigma_z} \cG(x,z) \, \mu_{I_1}^{J_1}(z) \, V_{ \, I_2 \cdots \, I_n \, I }^{J_2 \cdots J_n \, J}  (z,y) 
\no \\
W_{I_1 \cdots I_n \, I \, L} ^{J_1 \cdots J_n \, J}  (x,y) & = & 
{ i \over 2} \int_{\Sigma_z} W_{I_1} (x,z) \oom^{J_1} (z) W_{I_2 \cdots I_n \, I \, L} ^{J_2 \cdots J_n \, J}  (z,y) 
\eea
and that the modular graph tensors admit alternative definitions,
\begin{align}
\cA_{I_1 \cdots I_n} ^{J_1 \cdots J_n} &= \int_{\Sigma^2_{x,y}} \mu_{I_1}^{J_1}(x) V_{I_2\ldots I_{n-1}}^{J_2\ldots J_{n-1}}(x,y) \mu_{I_n}^{J_n}(y)
\label{altAdef} \\
&= \int_{\Sigma_x} \Phi^{J_{k-1} \ldots J_2 J_1}_{\, I_{k-1} \ldots I_2 I_1}(x) \mu^{J_k}_{I_k}(x) \Phi^{J_{k+1}\ldots J_n}_{\,I_{k+1} \ldots I_n}(x)  \notag
\end{align}
where one can choose any of $k=1,2,\ldots,n$ in the second line.
Moreover, we introduce complex conjugate versions of the $W$ tensors in (\ref{lemma1b}) and (\ref{VWn})
\begin{align}
\overline{W}^{I}(x,y) &= \cG(x,y) \overline \omega^I(y) - \Phi^I_J(x) \overline \omega^J(y)
\label{ccWs} \\
\overline{W}^{I_1 \ldots I_n L}_{J_1\ldots J_n}(x,y) &= \frac{i}{2} \int_{\Sigma_z}
\overline{W}^{I_1 \ldots I_{n-1}I_n}_{J_1\ldots J_{n-1}}(x,z) \omega_{J_n}(z) 
\overline{W}^{L}(z,y) 
\notag
\end{align}
which for instance yield the following complex conjugate of (\ref{otherW2}):
\bea
\overline{W}^{I L}_{J}(x,y) = \Big( V^I_J(x,y) - \Phi^I_\alpha(x) \Phi^\alpha_J(y) \Big) \overline \omega^L(y)
- \Big( \Phi^{IL}_{J M}(x) - \Phi^I_\alpha(x) \cA^{\alpha L}_{JM} \Big) \overline \omega^M(y)
\eea

\subsection{Graphical representation}
\label{sec:graph}

A natural graphical representation may be formulated for the tensors $V^{J_1 \cdots J_n} _{\, I_1 \cdots \, I_n} (x,y)$, $\Phi ^{J_1 \cdots J_n} _{I_1 \cdots I_n} (x)$ and $\cA ^{J_1 \cdots J_n} _{I_1 \cdots I_n}$ by representing each integrated vertex point $z_i$ by a black dot, each unintegrated vertex point by a white dot, and each Green function by a full line between two vertex points. In addition the integrated vertices $z_i$ carry one upper and one lower index, corresponding to the measure of integration provided by the $(1,1)$-differential $\mu^J_I(z_i)$ at the vertex $z_i$.
\begin{center}
\tikzpicture[scale=1]
\scope[xshift=0cm,yshift=0cm]
\draw (-1.7,0) node{$ V^{J_1 \cdots J_n} _{\,I_1 \cdots I_n} (x,y) =$};
\draw[very thick] (0,0) -- (8,0);
\draw[fill=white] (0,0)  circle [radius=.07] ;
\draw[fill=black] (1,0)  circle [radius=.07] ;
\draw[fill=black] (2,0)  circle [radius=.07] ;
\draw[fill=black] (6,0)  circle [radius=.07] ;
\draw[fill=black] (7,0)  circle [radius=.07] ;
\draw[fill=white] (8,0)  circle [radius=.07] ;
\draw (0,0.4) node{$x$};
\draw (8,0.4) node{$y$};
\draw (1,-0.4) node{\small $I_1$};
\draw (1,0.4) node{\small $J_1$};
\draw (2,-0.4) node{\small $I_2$};
\draw (2,0.4) node{\small $J_2$};
\draw (6,-0.4) node{\small $I_{n-1}$};
\draw (6,0.4) node{\small $J_{n-1}$};
\draw (7,-0.4) node{\small $I_n$};
\draw (7,0.4) node{\small $J_n$};
\draw (4, 0.3) node{$\cdots$};
\draw (4, -0.3) node{$\cdots$};
\endscope
\scope[xshift=8cm,yshift=0cm]
\draw (0.5, 0) node{$=$};
\draw[very thick] (1,0) -- (1.5,0);
\draw[very thick] (3.5,0) -- (4,0);
\draw[thick] (1.5,-0.5) -- (3.5,-0.5);
\draw[thick] (1.5,0.5) -- (3.5,0.5);
\draw[thick] (1.5,-0.5) -- (1.5,0.5);
\draw[thick] (3.5,-0.5) -- (3.5,0.5);
\draw[fill=white] (1,0)  circle [radius=.07] ;
\draw (1,0.4) node{$x$};
\draw[fill=white] (4,0)  circle [radius=.07] ;
\draw (4,0.4) node{$y$};
\draw (2.5,0) node{$ V^{J_1 \cdots J_n} _{\,I_1 \cdots I_n}$};
\endscope
\scope[xshift=0cm,yshift=-1.8cm]
\draw (-1.5,0) node{$ \Phi^{J_1 \cdots J_n} _{\,I_1 \cdots I_n} (x) =$};
\draw[very thick] (0,0) -- (7,0);
\draw[fill=white] (0,0)  circle [radius=.07] ;
\draw[fill=black] (1,0)  circle [radius=.07] ;
\draw[fill=black] (2,0)  circle [radius=.07] ;
\draw[fill=black] (6,0)  circle [radius=.07] ;
\draw[fill=black] (7,0)  circle [radius=.07] ;
\draw (0,0.4) node{$x$};
\draw (1,-0.4) node{\small $I_1$};
\draw (1,0.4) node{\small $J_1$};
\draw (2,-0.4) node{\small $I_2$};
\draw (2,0.4) node{\small $J_2$};
\draw (6,-0.4) node{\small $I_{n-1}$};
\draw (6,0.4) node{\small $J_{n-1}$};
\draw (7,-0.4) node{\small $I_n$};
\draw (7,0.4) node{\small $J_n$};
\draw (4, 0.3) node{$\cdots$};
\draw (4, -0.3) node{$\cdots$};
\endscope
\scope[xshift=8cm,yshift=-1.8cm]
\draw (0.5, 0) node{$=$};
\draw[very thick] (1,0) -- (1.5,0);
\draw[thick] (1.5,-0.5) -- (3.5,-0.5);
\draw[thick] (1.5,0.5) -- (3.5,0.5);
\draw[thick] (1.5,-0.5) -- (1.5,0.5);
\draw[thick] (3.5,-0.5) -- (3.5,0.5);
\draw[fill=white] (1,0)  circle [radius=.07] ;
\draw (1,0.4) node{$x$};
\draw (2.5,0) node{$ \Phi^{J_1 \cdots J_n} _{\, I_1 \cdots I_n}$};
\endscope
\scope[xshift=0cm,yshift=-3.6cm]
\draw (-1.2,0) node{$ \cA^{J_1 \cdots J_n} _{\, I_1 \cdots I_n}  =$};
\draw[very thick] (1,0) -- (7,0);
\draw[fill=black] (1,0)  circle [radius=.07] ;
\draw[fill=black] (2,0)  circle [radius=.07] ;
\draw[fill=black] (6,0)  circle [radius=.07] ;
\draw[fill=black] (7,0)  circle [radius=.07] ;
\draw (1,-0.4) node{\small $I_1$};
\draw (1,0.4) node{\small $J_1$};
\draw (2,-0.4) node{\small $I_2$};
\draw (2,0.4) node{\small $J_2$};
\draw (6,-0.4) node{\small $I_{n-1}$};
\draw (6,0.4) node{\small $J_{n-1}$};
\draw (7,-0.4) node{\small $I_n$};
\draw (7,0.4) node{\small $J_n$};
\draw (4, 0.3) node{$\cdots$};
\draw (4, -0.3) node{$\cdots$};
\endscope
\scope[xshift=8cm,yshift=-3.6cm]
\draw (0.5, 0) node{$=$};
\draw[thick] (1.5,-0.5) -- (3.5,-0.5);
\draw[thick] (1.5,0.5) -- (3.5,0.5);
\draw[thick] (1.5,-0.5) -- (1.5,0.5);
\draw[thick] (3.5,-0.5) -- (3.5,0.5);
\draw (2.5,0) node{$ \cA^{J_1 \cdots J_n} _{\, I_1 \cdots I_n}$};
\endscope
\endtikzpicture
\end{center}
Using this graphical representation for the tensors $V^{J_1 \cdots J_n} _{I_1 \cdots I_n} (x,y) $, $\Phi^{J_1 \cdots J_n} _{I_1 \cdots I_n} (x)$, and $\cA^{J_1 \cdots J_n} _{I_1 \cdots I_n}$, we may express the recursion relations of (\ref{VWn}) amongst them   as follows,
\begin{center}
\tikzpicture[scale=1]
\scope[xshift=0cm,yshift=0cm]
\draw[very thick] (1,0) -- (1.5,0);
\draw[very thick] (3.5,0) -- (4,0);
\draw[thick] (1.5,-0.5) -- (3.5,-0.5);
\draw[thick] (1.5,0.5) -- (3.5,0.5);
\draw[thick] (1.5,-0.5) -- (1.5,0.5);
\draw[thick] (3.5,-0.5) -- (3.5,0.5);
\draw[fill=white] (1,0)  circle [radius=.07] ;
\draw (1,0.4) node{$x$};
\draw[fill=white] (4,0)  circle [radius=.07] ;
\draw (4,0.4) node{$y$};
\draw (2.5,0) node{$ V^{J_1 \cdots J_n \, J} _{\, I_1 \cdots I_n \, I}$};
\endscope
\scope[xshift=4.5cm,yshift=0cm]
\draw (0.3, 0) node{$=$};
\draw[fill=black] (4,0)  circle [radius=.07] ;
\draw[very thick] (4,0) -- (5,0);
\draw (4,-0.4) node{\small $I$};
\draw (4,0.4) node{\small $J$};
\draw[very thick] (1,0) -- (1.5,0);
\draw[very thick] (3.5,0) -- (4,0);
\draw[thick] (1.5,-0.5) -- (3.5,-0.5);
\draw[thick] (1.5,0.5) -- (3.5,0.5);
\draw[thick] (1.5,-0.5) -- (1.5,0.5);
\draw[thick] (3.5,-0.5) -- (3.5,0.5);
\draw[fill=white] (1,0)  circle [radius=.07] ;
\draw (1,0.4) node{$x$};
\draw[fill=white] (5,0)  circle [radius=.07] ;
\draw (5,0.4) node{$y$};
\draw (2.5,0) node{$ V^{J_1 \cdots J_n} _{\,I_1 \cdots I_n}$};
\endscope
\scope[xshift=0cm,yshift=-1.8cm]
\draw[very thick] (1,0) -- (1.5,0);
\draw[thick] (1.5,-0.5) -- (3.5,-0.5);
\draw[thick] (1.5,0.5) -- (3.5,0.5);
\draw[thick] (1.5,-0.5) -- (1.5,0.5);
\draw[thick] (3.5,-0.5) -- (3.5,0.5);
\draw[fill=white] (1,0)  circle [radius=.07] ;
\draw (1,0.4) node{$x$};
\draw (2.5,0) node{$ \Phi^{J \, J_1 \cdots J_n} _{I \, I_1 \cdots I_n}$};
\endscope
\scope[xshift=4.5cm,yshift=-1.8cm]
\draw (0.3, 0) node{$=$};
\draw[fill=black] (2,0)  circle [radius=.07] ;
\draw (2,-0.4) node{\small $I$};
\draw (2,0.4) node{\small $J$};
\draw[very thick] (1,0) -- (2,0);
\draw[very thick] (2,0) -- (2.5,0);
\draw[thick] (2.5,-0.5) -- (4.5,-0.5);
\draw[thick] (2.5,0.5) -- (4.5,0.5);
\draw[thick] (2.5,-0.5) -- (2.5,0.5);
\draw[thick] (4.5,-0.5) -- (4.5,0.5);
\draw[fill=white] (1,0)  circle [radius=.07] ;
\draw (1,0.4) node{$x$};
\draw (3.5,0) node{$ \Phi^{J_1 \cdots J_n} _{I_1 \cdots I_n}$};
\endscope
\scope[xshift=0cm,yshift=-3.6cm]
\draw[thick] (1.5,-0.5) -- (3.5,-0.5);
\draw[thick] (1.5,0.5) -- (3.5,0.5);
\draw[thick] (1.5,-0.5) -- (1.5,0.5);
\draw[thick] (3.5,-0.5) -- (3.5,0.5);
\draw (2.5,0) node{$ \cA^{J \, J_1 \cdots J_n} _{I \, I_1 \cdots I_n}$};
\endscope
\scope[xshift=4.5cm,yshift=-3.6cm]
\draw (0.5, 0) node{$=$};
\draw[fill=black] (1,0)  circle [radius=.07] ;
\draw (1,-0.4) node{\small $I$};
\draw (1,0.4) node{\small $J$};
\draw[very thick] (1,0) -- (1.5,0);
\draw[thick] (1.5,-0.5) -- (3.5,-0.5);
\draw[thick] (1.5,0.5) -- (3.5,0.5);
\draw[thick] (1.5,-0.5) -- (1.5,0.5);
\draw[thick] (3.5,-0.5) -- (3.5,0.5);
\draw (2.5,0) node{$ \Phi^{J_1 \cdots J_n} _{I_1 \cdots I_n}$};
\endscope
\endtikzpicture
\end{center}
The recursion relation for the tensors $W_{I_1 \cdots I_n}^{J_1 \cdots J_n}(x,y)$ given in (\ref{VWn}) is not manifestly obtained by integrating against the differential forms $\mu_I^J(z_i)$, and we shall postpone its graphical representation until the next subsection.

\subsection{Decomposition and recursion relations of the $W$ tensors}

In this subsection, we obtain the expressions for $W$, defined recursively in the second line of 
(\ref{VWn}), in terms of the building blocks $V$, $\Phi$ and $\cA$. These expressions will be useful later in constructing explicit forms for the identities among modular graph tensors. Given the general structure of $W$, it is convenient to decompose it as follows,
\bea
W^{J_1 \cdots J_n} _{I_1 \cdots I_n L} (x,y) 
= C^{J_1 \cdots J_n} _{I_1 \cdots I_n } (x,y) \, \om_L(y) - D^{J_1 \cdots J_n K} _{I_1 \cdots I_n \, L} (x) \, \om_K(y)
\label{CDS.1}
\eea
Substituting this decomposition into the recursion relation for $W$ in (\ref{VW}), and decomposing the resulting relation in terms of the tensors $C$ and $D$, we obtain the following coupled recursion relations for the tensors $C$ and $D$,\footnote{We are deeply grateful to one of the referees for suggesting to express the recursion relations for $W$ directly in terms of recursion relations for $C$ and $D$.}
\bea
C_{\, I_1 \cdots I_n \, I} ^{J_1 \cdots J_n \, J} (x,y)
& = & 
\int_{\Sigma _z} C_{\, I_1 \cdots I_n } ^{J_1 \cdots J_n } (x,z) \mu_I^J(z) \cG(z,y)
- D_{\, I_1 \cdots I_n \, I} ^{J_1 \cdots J_n \, \a} (x) \, \Phi _\a ^J (y)
\no \\
D_{\, I_1 \cdots I_n \, IL} ^{J_1 \cdots J_n \, JK} (x)
& = & 
\int_{\Sigma _z} C_{\, I_1 \cdots I_n } ^{J_1 \cdots J_n } (x,z) \mu_I^J(z) \Phi _L^K(z)
- D_{\, I_1 \cdots I_n \, I} ^{J_1 \cdots J_n \, \a} (x) \, \cA _{\a L} ^{JK}
\label{CDrecursion}
\eea
The graphical representation of these recursion relations takes on the following from,
\begin{center}
\tikzpicture[scale=1]
\scope[xshift=0.1cm,yshift=0cm]
\draw[very thick] (1,0) -- (1.5,0);
\draw[very thick] (3.5,0) -- (4,0);
\draw[fill=white] (1,0)  circle [radius=.07] ;
\draw (1,0.4) node{$x$};
\draw[fill=white] (4,0)  circle [radius=.07] ;
\draw (4,0.4) node{$y$};
\draw[thick] (1.5,-0.5) -- (3.5,-0.5);
\draw[thick] (1.5,0.5) -- (3.5,0.5);
\draw[thick] (1.5,-0.5) -- (1.5,0.5);
\draw[thick] (3.5,-0.5) -- (3.5,0.5);
\draw (2.5,0) node{$ C^{J_1 \cdots J_n \, J} _{\, I_1 \cdots I_n \, I}$};
\endscope
\scope[xshift=4.5cm,yshift=0cm]
\draw (0.3, 0) node{$=$};
\draw[fill=black] (4,0)  circle [radius=.07] ;
\draw[very thick] (4,0) -- (5,0);
\draw[fill=white] (5,0)  circle [radius=.07] ;
\draw (5,0.4) node{$y$};
\draw (4,-0.4) node{\small $I$};
\draw (4,0.4) node{\small $J$};
\draw[very thick] (1,0) -- (1.5,0);
\draw[very thick] (3.5,0) -- (4,0);
\draw[thick] (1.5,-0.5) -- (3.5,-0.5);
\draw[thick] (1.5,0.5) -- (3.5,0.5);
\draw[thick] (1.5,-0.5) -- (1.5,0.5);
\draw[thick] (3.5,-0.5) -- (3.5,0.5);
\draw[fill=white] (1,0)  circle [radius=.07] ;
\draw (1,0.4) node{$x$};
\draw (2.5,0) node{$ C^{J_1 \cdots J_n} _{\, I_1 \cdots I_n}$};
\endscope
\scope[xshift=9.5cm,yshift=0cm]
\draw (0.47, 0) node{$-$};
\draw[very thick] (1,0) -- (1.5,0);
\draw[fill=white] (1,0)  circle [radius=.07] ;
\draw (1,0.4) node{$x$};
\draw[thick] (1.5,-0.5) -- (3.5,-0.5);
\draw[thick] (1.5,0.5) -- (3.5,0.5);
\draw[thick] (1.5,-0.5) -- (1.5,0.5);
\draw[thick] (3.5,-0.5) -- (3.5,0.5);
\draw (2.5,0) node{$ D^{J_1 \cdots J_n\, \a} _{\, I_1 \cdots I_n \, I}$};
\endscope
\scope[xshift=12.5cm,yshift=0cm]
\draw (1, 0) node{$\times$};
\draw[very thick] (2.5,0) -- (3,0);
\draw[fill=white] (3,0)  circle [radius=.07] ;
\draw (3,0.4) node{$y$};
\draw[thick] (1.5,-0.5) -- (2.5,-0.5);
\draw[thick] (1.5,0.5) -- (2.5,0.5);
\draw[thick] (1.5,-0.5) -- (1.5,0.5);
\draw[thick] (2.5,-0.5) -- (2.5,0.5);
\draw (2,0) node{$ \Phi^{J} _\a$};
\endscope
\scope[xshift=0.1cm,yshift=-1.8cm]
\draw[very thick] (1,0) -- (1.5,0);
\draw[fill=white] (1,0)  circle [radius=.07] ;
\draw (1,0.4) node{$x$};
\draw[thick] (1.5,-0.5) -- (3.5,-0.5);
\draw[thick] (1.5,0.5) -- (3.5,0.5);
\draw[thick] (1.5,-0.5) -- (1.5,0.5);
\draw[thick] (3.5,-0.5) -- (3.5,0.5);
\draw (2.5,0) node{$ D^{J_1 \cdots J_n\, JK} _{\, I_1 \cdots I_n \, IL}$};
\endscope
\scope[xshift=4.5cm,yshift=-1.8cm]
\draw (0.3, 0) node{$=$};
\draw[very thick] (1,0) -- (1.5,0);
\draw[fill=white] (1,0)  circle [radius=.07] ;
\draw (1,0.4) node{$x$};
\draw[thick] (1.5,-0.5) -- (3.5,-0.5);
\draw[thick] (1.5,0.5) -- (3.5,0.5);
\draw[thick] (1.5,-0.5) -- (1.5,0.5);
\draw[thick] (3.5,-0.5) -- (3.5,0.5);
\draw (2.5,0) node{$ C^{J_1 \cdots J_n} _{\, I_1 \cdots I_n}$};
\draw[very thick] (3.5,0) -- (4,0);
\draw[fill=black] (4,0)  circle [radius=.07] ;
\draw (4,-0.4) node{\small $I$};
\draw (4,0.4) node{\small $J$};
\endscope
\scope[xshift=7.5cm,yshift=-1.8cm]
\draw[very thick] (1,0) -- (1.5,0);
\draw[thick] (1.5,-0.5) -- (2.5,-0.5);
\draw[thick] (1.5,0.5) -- (2.5,0.5);
\draw[thick] (1.5,-0.5) -- (1.5,0.5);
\draw[thick] (2.5,-0.5) -- (2.5,0.5);
\draw (2,0) node{$ \Phi^K_L$};
\endscope
\scope[xshift=10cm,yshift=-1.8cm]
\draw (0.47, 0) node{$-$};
\draw[very thick] (1,0) -- (1.5,0);
\draw[fill=white] (1,0)  circle [radius=.07] ;
\draw (1,0.4) node{$x$};
\draw[thick] (1.5,-0.5) -- (3.5,-0.5);
\draw[thick] (1.5,0.5) -- (3.5,0.5);
\draw[thick] (1.5,-0.5) -- (1.5,0.5);
\draw[thick] (3.5,-0.5) -- (3.5,0.5);
\draw (2.5,0) node{$ D^{J_1 \cdots J_n\, \a} _{\, I_1 \cdots I_n \, I}$};
\endscope
\scope[xshift=13cm,yshift=-1.8cm]
\draw (1, 0) node{$\times$};
\draw[thick] (1.4,-0.5) -- (2.6,-0.5);
\draw[thick] (1.4,0.5) -- (2.6,0.5);
\draw[thick] (1.4,-0.5) -- (1.4,0.5);
\draw[thick] (2.6,-0.5) -- (2.6,0.5);
\draw (2,0) node{$ \cA^{JK} _{\a L}$};
\endscope
\endtikzpicture
\end{center}
The simplest examples of $C,D$ can be read off from the expressions in (\ref{lemma1b}) and (\ref{otherW2}),
\begin{align}
C(x,y) &= \cG(x,y) \, , &C^{J_1}_{I_1}(x,y) &= V^{J_1}_{I_1}(x,y) - \Phi^{\a}_{I_1}(x) \Phi^{J_1}_{\a}(y)  \notag\\
D^K_L(x) &= \Phi^K_L(x) \, , &D^{J_1 K}_{I_1 L}(x) &= \Phi^{J_1 K}_{I_1 L}(x) - \Phi^{\a}_{I_1}(x) \cA^{J_1K}_{\, \a\, L} \label{CDS.2}
\end{align}
For weight 3 we obtain, 
\bea
C^{J_1 J_2} _{I_1 I_2 } (x,y) & = & 
V^{J_1 J_2} _{I_1 I_2 } (x,y) 
- \Phi ^{\a} _{I_1} (x) \Phi^{J_2 J_1}_{I_2\, \a} (y) 
- \Phi^{J_1\a}_{I_1 I_2}(x) \Phi_{\a}^{J_2}(y) 
+ \Phi^{\a}_{I_1}(x) \cA^{J_1\b }_{\a I_2} \Phi _{\b}^{J_2}(y) 
\no \\ 
D^{J_1 J_2 K} _{I_1 I_2 \, L} (x) & = & 
\Phi ^{J_1 J_2 K} _{I_1 I_2 L} (x) 
- \Phi ^{J_1 \a} _{I_1 I_2} (x) \cA^{J_2 K}_{\a L} 
- \Phi^{\a}_{I_1}(x) \cA^{J_1 J_2 K}_{\a I_2 L} 
+ \Phi^{\a}_{I_1}(x) \cA^{J_1\b }_{\a I_2} \cA_{\b L}^{J_2 K} 
\label{CDS.3}
\eea
For weight 4, we obtain, 
\bea
C^{J_1 J_2J_3} _{I_1 I_2 I_3 } (x,y) & = &
V^{J_1 J_2J_3} _{I_1 I_2 I_3 } (x,y) 
- \Phi _{I_1 I_2} ^{J_1 \a} (x) \, \Phi_{I_3 \, \a} ^{J_3 J_2}(y)
- \Phi _{I_1 I_2 I_3} ^{J_1 J_2 \, \a} (x) \, \Phi_{ \a} ^{J_3 }(y)
+\Phi _{I_1I_2}^{J_1\a} (x) \, \cA_{\a I_3} ^{J_2 \b} \, \Phi_\b ^{J_3}(y)
\no \\ &&
- \Phi _{I_1}^\a (x) \left \{ 
\Phi ^{J_3 J_2J_1} _{I_3 I_2 \, \a } (y) 
- \cA _{\a I_2} ^{J_1 \b} \, \Phi_{I_3 \, \b} ^{J_3 J_2}(y)
- \cA _{\a I_2 I_3} ^{J_1 J_2 \, \b} \, \Phi_{ \b} ^{J_3 }(y)
+ \cA _{ \a I_2}^{J_1\b} \, \cA_{\b I_3} ^{J_2  \g} \, \Phi_\g ^{J_3}(y)
\right \}
\no \\
D^{J_1 J_2 J_3 K} _{I_1 I_2 I_3 \, L} (x) & = &
\Phi_{I_1I_2I_3 \, L} ^{J_1J_2J_3K}(x)
- \Phi_{I_1 I_2 I_3}^{J_1 J_2 \a} (x) \, \cA_{\a L} ^{J_3 K}
- \Phi_{I_1 I_2}^{J_1  \a} (x) \, \cA_{\a I_3 L} ^{J_2 J_3 K}
+\Phi_{I_1 I_2}^{J_1  \a} (x) \, \cA_{\a I_3} ^{J_2 \b } \, \cA _{\b L}^{J_3K}
\no \\ &&
- \Phi _{I_1}^\a (x) \left \{ 
\cA_{ \a I_2I_3L} ^{J_1J_2J_3K}
- \cA_{\a  I_2 I_3}^{J_1 J_2 \, \b} \, \cA_{\b L} ^{J_3 K}
- \cA_{\a I_2}^{J_1  \b} \, \cA_{\b I_3 L} ^{J_2 J_3 K}
+\cA_{\a I_2}^{J_1  \b} \, \cA_{\b I_3} ^{J_2 \g } \, \cA _{\g L}^{J_3K}
\right \}
\label{CDS.4}
\qquad
\eea
At higher weight $n$, we obtain
\begin{align}
C^{J_1 J_2 \dots J_n}_{\, I_1 I_2 \,\ldots \, I_n}(x,y) &= V^{J_1 J_2 \dots J_n}_{\, I_1 I_2\, \ldots \, I_n}(x,y) + \sum_{k=1}^n (-1)^k
\sum_{1\leq i_1<i_2<\ldots <i_k \leq n}
\Phi^{J_1  \ldots J_{i_1-1} \alpha_1}_{ I_1\, \ldots I_{i_1-1} I_{i_1}}(x)
\cA^{J_{i_1} J_{i_1+1} \ldots J_{i_2-1} \alpha_2}_{\, \alpha_1 \, I_{i_1+1} \ldots I_{i_2-1} I_{i_2} }
\notag \\
& \ \  \ \    \times \cA^{J_{i_2} J_{i_2+1} \ldots J_{i_3-1} \alpha_3}_{\, \alpha_2 \, I_{i_2+1} \ldots I_{i_3-1} I_{i_3} }  \ldots 
\cA^{J_{i_{k-1}} J_{i_{k-1}+1} \ldots J_{i_k-1} \alpha_k}_{\, \alpha_{k-1} \, I_{i_{k-1}+1} \ldots I_{i_k-1} I_{i_k} } 
\Phi^{J_n  J_{n-1} \ldots  J_{i_k+1} J_{i_k} }_{ I_n \, I_{n-1} \ldots \,  I_{i_k+1}  \, \alpha_k \,} (y) \label{CDS.5}\\
D^{J_1 J_2 \dots J_n K}_{\,I_1 I_2 \, \ldots \, I_n L}(x) &= \Phi^{J_1 J_2 \dots J_n K}_{ \, I_1 I_2\,  \ldots \, I_n L}(x) + \sum_{k=1}^n (-1)^k
\sum_{1\leq i_1<i_2<\ldots <i_k \leq n}
\Phi^{J_1  \ldots J_{i_1-1} \alpha_1}_{ I_1\, \ldots I_{i_1-1} I_{i_1}}(x)
\cA^{J_{i_1} J_{i_1+1} \ldots J_{i_2-1} \alpha_2}_{\, \alpha_1 \, I_{i_1+1} \ldots I_{i_2-1} I_{i_2} }
\notag \\
& \ \  \ \    \times \cA^{J_{i_2} J_{i_2+1} \ldots J_{i_3-1} \alpha_3}_{\, \alpha_2 \, I_{i_2+1} \ldots I_{i_3-1} I_{i_3} }  \ldots 
\cA^{J_{i_{k-1}} J_{i_{k-1}+1} \ldots J_{i_k-1} \alpha_k}_{\, \alpha_{k-1} \, I_{i_{k-1}+1} \ldots I_{i_k-1} I_{i_k} } 
\cA^{J_{i_k} J_{i_k+1} \ldots J_n K}_{\, \alpha_k \, I_{i_k+1} \ldots \, I_n  L}
\label{CDS.6}
\end{align}
which will be proven in appendix \ref{app:A}.
Together with the first terms $V^{J_1 J_2 \dots J_n}_{\, I_1 I_2\, \ldots \, I_n}(x,y)$
and $\Phi^{J_1 J_2 \dots J_n K}_{ \, I_1 I_2\,  \ldots \, I_n L}(x) $ on the right-hand sides
of (\ref{CDS.5}) and (\ref{CDS.6}), the respective sums over $\sum_{k=1}^n$ and $\sum_{1\leq i_1<i_2<\ldots <i_k \leq n}$ yield all the $2^n$ possibilities
to replace a subset of the $n$ pairs $\begin{smallmatrix} J_i \\ I_i \end{smallmatrix}$ by contractions like
$\cA^{\ldots \alpha}_{\ldots I_i} \cA_{\alpha \ldots }^{J_i \ldots}$ (or with ${\cal A}$ replaced by $\Phi$). 
Terms with an odd number of $\cA$-factors enter (\ref{CDS.5}) with a plus sign and (\ref{CDS.6}) with a minus sign.

Note that the contributions to the $k$-summands in $D^{J_1 J_2 \dots J_n K}_{\,I_1 I_2 \, \ldots \, I_n L}(x)$ 
are formally obtained from those to $C^{J_1 J_2 \dots J_n}_{\, I_1 I_2 \,\ldots \, I_n}(x,y)$
by replacing the last factors $\Phi^{J_n  J_{n-1} \ldots  J_{i_k+1} J_{i_k} }_{ I_n \, I_{n-1} \ldots \,  I_{i_k+1}  \, \alpha_k \,} (y)  \rightarrow \cA^{J_{i_k} J_{i_k+1} \ldots J_n K}_{\, \alpha_k \, I_{i_k+1} \ldots \, I_n  L}$.
Hence, one can equivalently write
\bea
W^{J_1 \cdots J_n} _{I_1 \cdots I_n L} (x,y)  &=&  
V^{J_1 J_2 \dots J_n}_{\, I_1 I_2\, \ldots \, I_n}(x,y)\omega_L(y)
- \Phi^{J_1 J_2 \dots J_n K}_{ \, I_1 I_2\,  \ldots \, I_n L}(x) \omega_K(y)  
\no \\ &&
+ \sum_{k=1}^n (-1)^k \!\!\!
\sum_{1\leq i_1<i_2<\ldots <i_k \leq n} \!\!\!
\Phi^{J_1  \ldots J_{i_1-1} \alpha_1}_{ I_1\, \ldots I_{i_1-1} I_{i_1}}(x) \, 
\cA^{J_{i_1} J_{i_1+1} \ldots J_{i_2-1} \alpha_2}_{\, \alpha_1 \, I_{i_1+1} \ldots I_{i_2-1} I_{i_2} }
\no \\ && \qquad  \qquad \times 
\cA^{J_{i_2} J_{i_2+1} \ldots J_{i_3-1} \alpha_3}_{\, \alpha_2 \, I_{i_2+1} \ldots I_{i_3-1} I_{i_3} }  
~ \times \cdots \times ~
\cA^{J_{i_{k-1}} J_{i_{k-1}+1} \ldots J_{i_k-1} \alpha_k}_{\, \alpha_{k-1} \, I_{i_{k-1}+1} \ldots I_{i_k-1} I_{i_k} } 
\no \\ && \qquad \qquad \times 
\Big (  
\Phi^{J_n  J_{n-1} \ldots  J_{i_k+1} J_{i_k} }_{ I_n \, I_{n-1} \ldots \,  I_{i_k+1}  \, \alpha_k \,} (y) \omega_L(y) 
- \cA^{J_{i_k} J_{i_k+1} \ldots J_n K}_{\, \alpha_k \, I_{i_k+1} \ldots \, I_n  L}\omega_K(y) \Big )
\label{CDS.7}
\eea
The analogous results for the complex conjugate versions (\ref{ccWs}) of the $W$ tensors 
involve contractions of the form $\cA_{\ldots \alpha}^{\ldots I_i} \cA^{\alpha \ldots }_{J_i \ldots}$
instead of $\cA^{\ldots \alpha}_{\ldots I_i} \cA_{\alpha \ldots }^{J_i \ldots}$
\bea
\overline{W}_{J_1 \cdots J_n}^{I_1 \cdots I_n L} (x,y)  &=&  
V_{J_1 J_2 \dots J_n}^{\, I_1 I_2\, \ldots \, I_n}(x,y) \overline{\omega}^L(y)
- \Phi_{J_1 J_2 \dots J_n K}^{ \, I_1 I_2\,  \ldots \, I_n L}(x) \overline{\omega}^K(y)  
\no \\ &&
+ \sum_{k=1}^n (-1)^k \!\!\!
\sum_{1\leq i_1<i_2<\ldots <i_k \leq n} \!\!\!
\Phi_{J_1  \ldots J_{i_1-1} \alpha_1}^{ I_1\, \ldots I_{i_1-1} I_{i_1}}(x) \, 
\cA_{J_{i_1} J_{i_1+1} \ldots J_{i_2-1} \alpha_2}^{\, \alpha_1 \, I_{i_1+1} \ldots I_{i_2-1} I_{i_2} }
\no \\ && \qquad  \qquad \times 
\cA_{J_{i_2} J_{i_2+1} \ldots J_{i_3-1} \alpha_3}^{\, \alpha_2 \, I_{i_2+1} \ldots I_{i_3-1} I_{i_3} }  
~ \times \cdots \times ~
\cA_{J_{i_{k-1}} J_{i_{k-1}+1} \ldots J_{i_k-1} \alpha_k}^{\, \alpha_{k-1} \, I_{i_{k-1}+1} \ldots I_{i_k-1} I_{i_k} } 
\no \\ && \qquad \qquad \times 
\Big (  
\Phi_{J_n  J_{n-1} \ldots  J_{i_k+1} J_{i_k} }^{ I_n \, I_{n-1} \ldots \,  I_{i_k+1}  \, \alpha_k \,} (y) \overline \omega^L(y) 
- \cA_{J_{i_k} J_{i_k+1} \ldots J_n K}^{\, \alpha_k \, I_{i_k+1} \ldots \, I_n  L} \overline \omega^K(y) \Big )
\label{CDS.8}
\eea

 \subsection{The generalized interchange lemma}
 
 We are now in a position to state and prove the generalization of Lemma \ref{lemma1} that will be at the root of the identities derived here.
 
 {\lem
\label{lemma2}
On a compact Riemann surface of  arbitrary genus $h \geq 1$, the following relation between derivatives of the tensor functions $W_{I_1 \cdots I_n L}^{J_1 \cdots J_n}(x,y)$ holds,
\bea
\label{lemma2a}
\p_x W_{I_1 \cdots I_n L}^{J_1 \cdots J_n}(x,y) = - \p_y W_{L I_n \cdots I_1 }^{J_n \cdots J_1}(y,x)
\eea
where the  tensor $W$ was defined recursively in (\ref{VW}) and (\ref{VWn}). }

\sm

 The proof of Lemma \ref{lemma2} proceeds by induction on $n$. It was proven for $n=0$ in Lemma \ref{lemma1}, and for $n=1$ in (\ref{W2}). Assuming that (\ref{lemma2a}) holds at step $n$, we shall now prove the corresponding relation for step $n+1$ in the induction, by taking the $x$-derivative of both sides of the second equation in (\ref{VWn}), 
 \bea
\p_x W_{I_1 \cdots I_n \, I_{n+1} \, L} ^{J_1 \cdots J_n \, J_{n+1}}  (x,y) & = & 
{ i \over 2} \int _{\Sigma_z} \p_x W_{I_1 \cdots I_n \, I_{n+1}} ^{J_1 \cdots J_n }  (x,z)  \, \oom^{J_{n+1}}(z) \, W_L(z,y)
\eea
Using the assumption that (\ref{lemma2a}) holds at step $n$, we transform the $x$-derivative into a $z$-derivative and integrate by parts in $z$,
 \bea
\p_x W_{I_1 \cdots I_n \, I_{n+1} \, L} ^{J_1 \cdots J_n \, J_{n+1}}  (x,y) & = & 
 { i \over 2} \int _{\Sigma_z} W_{I_{n+1}  \, I_n \cdots I_1} ^{J_n \cdots J_1 }  (z,x)  \, \oom^{J_{n+1}} (z) \, \p_z W_L(z,y)
\eea
Finally, we use (\ref{lemma1a}) of Lemma \ref{lemma1} to convert the $z$-derivate on $W_L(z,y)$ into a $y$-derivative,
and rearrange the different factors as follows,
 \bea
\p_x W_{I_1 \cdots I_n \, I_{n+1} \, L} ^{J_1 \cdots J_n \, J_{n+1}}  (x,y) & = & 
- { i \over 2} \int _{\Sigma_z} \, \p_y W_L(y,z) \, \oom^{J_{n+1}} (z)  \, W_{I_{n+1} \, I_n \cdots I_1} ^{J_n \cdots J_1 }  (z,x)  
\eea
so that it is clear that the right side gives the right side of (\ref{lemma2a}) for the inductive step $n+1$, thereby completing the proof of Lemma \ref{lemma2}.

Note that the complex conjugate versions (\ref{ccWs}) of the $W$ tensors
obey an analogous interchange lemma that can be used to swap antiholomorphic derivatives,
\beq
\partial_{\bar x}\overline{W}_{J_1 \cdots J_n}^{I_1 \cdots I_n L} (x,y) = - \partial_{\bar y} \overline{W}_{J_n \cdots J_1}^{L I_n \cdots I_1} (y,x)
\eeq
The inductive proof of Lemma \ref{lemma2} may be readily adapted to the complex conjugate case.

\newpage

\section{Identities between modular graph tensors}
\label{sec:3}
\setcounter{equation}{0}

The modular graph tensors $\cA^{J_1 \cdots J_n}_{I_1 \cdots I_n}$ introduced in the preceding section all correspond to tree-level graphs as they form linear chains of Arakelov Green functions. The identities we shall derive and prove below involve also one-loop modular graph tensors, defined as follows,
\bea
\cB ^{J_1 \cdots J_n } _{I_1 \cdots I_n} = 
 \prod_{i=1}^n  \int _{\Sigma_{z_i}}  \mu_{I_i} ^{J_i} (z_i) \, \cG(z_i, z_{i+1})
=
\int _{\Sigma_z} V_{\, I_1 \cdots \, I_{n-1}} ^{J_1 \cdots J_{n-1}} (z,z) \, \mu_{I_n}^{J_n}(z)
\label{defbb}
\eea
where  we cyclically identify $z_{n+1}=z_1$ in the first expression. A graphical representation of $\cB$ is presented in the figure below.
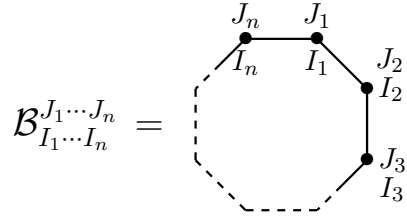
\begin{center}
\begin{tikzpicture} [scale=0.95, line width=0.30mm]
\draw (-2.7,-1.2) node{{\large $ \cB^{J_1 \cdots J_n} _{I_1 \cdots I_n} \,  =$}};
\draw (0.5,0)--(-0.5,0);
\draw (-0.5,0)--(-0.85,-0.35);
\draw [dashed](-0.85,-0.35)--(-1.2,-0.7);
\draw (0.5,0)--(1.2,-0.7);
\draw[dashed] (-1.2,-1.7)--(-1.2,-0.7);
\draw (1.2,-1.7)--(1.2,-0.7);
\draw (1.2,-1.7)--(0.85,-2.05);
\draw[dashed] (0.85,-2.05)--(0.5,-2.4);
\draw[dashed] (-0.5,-2.4)--(0.5,-2.4);
\draw[dashed] (-0.5,-2.4)--(-1.2,-1.7);
\draw(-0.5,0)node{$\bullet$}node[above]{\small $J_n$}node[below]{\small $I_n$};
\draw(0.5,0)node{$\bullet$}node[above]{\small $J_1$}node[below]{\small $I_1$};
%
\draw(1.2,-0.7)node{$\bullet$}node[right]{\small $I_2$};
\draw(1.2,-1.7)node{$\bullet$}node[right]{\small $J_3$};
\draw(1.52,-0.3)node{\small $J_2$};
\draw(1.55,-2.1)node{\small $I_3$};
\end{tikzpicture}
\end{center}
The tensor $\cB$ is clearly invariant under cyclic permutations of the pairs of indices $(I_i,J_i)$ and under reflection, schematically represented by the following identities,
\bea
 \cB^{J_1 \, J_2 \, \cdots \, J_{n-1} \, J_n}_{\, I_1\, I_2 \, \cdots \, I_{n-1} \, I_n}
 =  \cB^{J_2 \, \cdots \, J_{n-1} \, J_n \, J_1}_{\, I_2 \,  \cdots \, I_{n-1} \, I_n\,  I_1} 
 =  \cB^{J_n \, J_{n-1} \, \cdots \, J_2 \, J_1}_{\, I_n \, I_{n-1} \, \cdots \, I_2 \, I_1}
 \eea
The transformation law under $Sp(2h,\ZZ)$ of the tensor $\cB^{J_1 \cdots J_n}_{I_1 \cdots I_n}$  is identical to the transformation law given in (\ref{Trans}) for the tensor $ \cA^{J_1 \cdots J_n}_{I_1 \cdots I_n}$ corresponding
to a chain graph.

\subsection{The main theorem}
\label{sec:mainthm}

Instead of deriving identities directly for $\cA$ and $\cB$, the simplicity of the exchange Lemma \ref{lemma2} in terms of $W$ suggests that the simplest form of the identities is obtained rather in terms of expressions built out of $W$. These identities will contain higher genus modular graph tensors associated with one-loop and linear tree-level graphs.

{\thm
\label{thm0}
On a compact Riemann surface of  genus $h \geq 1$, the tensors defined by,
\bea
\cT^{J_1 \cdots J_n NM} _{~ I_1 \cdots I_n LK}
=  \frac{i}{2}  \int _{\Sigma^2_{x,y}} W_{I_1 \cdots I_n L} ^{J_1 \cdots J_n} (x,y) \Big ( \cG(x,y) \mu_K^M(x) 
- \mu _\a ^M(x) \Phi _K^\a(y) \Big ) \oom^N(y)
\label{defttens}
\eea
satisfy the following symmetry property,
\bea
\label{thm0.1}
\cT^{J_1 \cdots J_n NM} _{~ I_1 \cdots I_n LK}
=
\cT^{J_n J_{n-1} \cdots J_1 MN} _{~ L \ \, I_n \ \cdots \ I_2 I_1 K}
\eea
of the corresponding modular graph tensors.}

\sm

To prove Theorem \ref{thm0}, we evaluate the following integral,
\bea
-{ 1 \over 4 \pi} \int _{\Sigma ^3_{x,y,z}} \p_x \pbz \cG(x,z) \, \cG(y,z) \, W_{I_1 \cdots I_n L} ^{J_1 \cdots J_n} (x,y) \oom^M(x) \om_K(z) \oom ^N(y)
\label{thm0.11}
\eea
in two different ways. First, by using (\ref{Ara2}) for the mixed double derivative of the Arakelov Green function, we obtain the left-hand side of (\ref{thm0.1}). Second, integrating by parts in both $x$ and $\bar z$, using the generalized interchange lemma (\ref{lemma2a}) to convert $\partial_x$ into $\partial_y$ and then integrating by parts  in $y$ results in the following expression, 
\bea
-{ 1 \over 4 \pi} \int _{\Sigma ^3_{x,y,z}} \cG(x,z) \, \p_y \pbz  \cG(y,z) \,  W_{L I_n \cdots I_1} ^{J_n \cdots J_1} (x,y) \oom^M(x) \om_K(z) \oom ^N(y)
\label{thm0.12}
\eea
Swapping the integration variables $x$ and $y$, we recover the right side of (\ref{thm0.1}), which completes the proof of the Theorem. We note that complex conjugation using (\ref{ccWs}) gives rise to an equivalent version of the Theorem, stating that the tensors
\begin{align}
\overline{\cT}^{J_1 \cdots J_n LK} _{I_1 \cdots I_n NM}&= \frac{i}{2} \int_{\Sigma^2_{x,y}} \overline{W}^{I_1 \ldots I_n L}_{J_1\ldots J_n}(x,y) \omega_N(y) \Big( 
\cG(x,y) \mu_M^K(x) - \Phi^K_\alpha(y) \mu^\alpha_M(x)
\Big) 
\label{cctheorem} 
\end{align}
complex conjugate to (\ref{defttens})
obey the symmetry properties
\bea
\overline{\cT}^{J_1 \cdots J_n LK} _{I_1 \cdots I_n NM}=
\overline{\cT}^{\; L \; J_n \; \ldots \ J_2 J_1 K} _{I_n I_{n-1} \cdots I_1 M N}
\label{ccthm} 
\eea
complex conjugate to (\ref{thm0.1}).

\sm 

We note that in the graphical representations of section \ref{sec:graph}, the
${\cal T}$ tensor in (\ref{defttens}) can be brought into the following form
after inserting the decomposition (\ref{CDS.1}) of the $W$ tensor:
\begin{center}
\tikzpicture[scale=1]
\draw[thick] (-1.3,0.5) rectangle (-3.5,-0.5);
\draw(-2.4,0)node{${\cal T}^{J_1\ldots J_n NM}_{\; I_1 \ldots I_n\, L\,K}$};
\draw(-0.7,0)node{$=$};
\draw[thick] (1,0.5) rectangle (3,-0.5);
\draw(2,0)node{$C^{J_1\ldots J_n}_{\, I_1 \ldots I_n}$};
\draw[very thick] (1,0)--(0.5,0)node{$\bullet$};
\draw[very thick] (3,0)--(3.5,0)node{$\bullet$};
\draw(0.5,0.4)node{$M$};
\draw(0.5,-0.4)node{$K$};
\draw(3.5,0.4)node{$N$};
\draw(3.5,-0.4)node{$L$};
\draw[very thick](0.5,0)  ..controls (-0.5,-0.0) and (-0.5,-0.8)  ..  (0.5,-0.8);
\draw[very thick](3.5,0)  ..controls (4.5,-0.0) and (4.5,-0.8)  ..  (3.5,-0.8);
\draw[very thick](0.5,-0.8) -- (3.5,-0.8);
\draw(4.8,0)node{$-$};
\scope[xshift=5.6cm]
\draw[thick] (1,0.5) rectangle (3,-0.5);
\draw(2,0)node{$D^{J_1\ldots J_n \beta}_{\, I_1 \ldots I_n\, \! L}$};
\draw[very thick] (1,0)--(0.5,0)node{$\bullet$};
\draw (3.5,0)node{$\bullet$};
\draw(0.5,0.4)node{$M$};
\draw(0.5,-0.4)node{$K$};
\draw(3.5,0.4)node{$N$};
\draw(3.5,-0.4)node{$\beta$};
\draw[very thick](0.5,0)  ..controls (-0.5,-0.0) and (-0.5,-0.8)  ..  (0.5,-0.8);
\draw[very thick](3.5,0)  ..controls (4.5,-0.0) and (4.5,-0.8)  ..  (3.5,-0.8);
\draw[very thick](0.5,-0.8) -- (3.5,-0.8);
\endscope
\draw(-0.7,-2)node{$-$};
\draw(0,-2)node{$\bullet$};
\draw(0,-1.6)node{$M$};
\draw(0,-2.4)node{$\alpha$};
\draw[very thick](0,-2)--(0.5,-2);
\draw[thick] (0.5,-1.5) rectangle (2.5,-2.5);
\draw(1.5,-2)node{$C^{J_1\ldots J_n}_{\, I_1 \ldots I_n}$};
\draw[very thick](2.5,-2)--(3.5,-2);
\draw(3,-2)node{$\bullet$};
\draw(3,-1.6)node{$N$};
\draw(3,-2.4)node{$L$};
\draw[thick] (3.5,-1.5) rectangle (4.5,-2.5);
\draw(4,-2)node{$\Phi^\alpha_K$};
\draw(5.2,-2)node{$+$};
\scope[xshift=5.9cm]
\draw(0,-2)node{$\bullet$};
\draw(0,-1.6)node{$M$};
\draw(0,-2.4)node{$\alpha$};
\draw[very thick](0,-2)--(0.5,-2);
\draw[thick] (0.5,-1.5) rectangle (2.5,-2.5);
\draw(1.5,-2)node{$D^{J_1\ldots J_n \beta}_{\, I_1 \ldots I_n \, \! L}$};
\draw[very thick](3,-2)--(3.5,-2);
\draw(3,-2)node{$\bullet$};
\draw(3,-1.6)node{$N$};
\draw(3,-2.4)node{$\beta$};
\draw[thick] (3.5,-1.5) rectangle (4.5,-2.5);
\draw(4,-2)node{$\Phi^\alpha_K$};
\endscope
\endtikzpicture
\end{center}


\subsection{Weight 2}

For weight 2, namely $n=0$, the tensor in Theorem \ref{thm0} reduces to,
\begin{align}
\cT^{NM}_{\,LK}
&=  \int_{\Sigma^2_{x,y}}  
\Big ( \cG(x,y) \mu_K^M(x) 
- \mu _\a ^M(x) \Phi _K^\a(y) \Big )
\Big ( \cG(x,y) \mu_L^N(y) 
- \Phi^\b_L(x) \mu_\b^N(y) \Big )  \notag
\\
& =   \cB^{MN}_{KL}  - \cA_{\alpha KL}^{NM \alpha}
- \cA_{\alpha LK}^{MN \alpha}
+ \cA_{\alpha L}^{M \beta} \cA_{\beta K}^{N \alpha}  \label{ids.1}
\end{align}
These rearrangements are simple consequences of the recursive definitions of various tensors including (\ref{VWn}) and (\ref{defbb}). By Theorem \ref{thm0}, the tensor in (\ref{ids.1}) obeys the symmetry $\cT^{NM}_{\,LK}= \cT^{MN}_{\,LK}$, which leads to the eight-term identity,
\bea
\label{weight2}
\cB^{MN}_{KL} - \cB^{NM}_{KL} 
=  \cA_{\alpha KL}^{NM \alpha}
+ \cA_{\alpha LK}^{MN \alpha}
-\cA_{\alpha KL}^{MN \alpha}
- \cA_{\alpha LK}^{NM \alpha}
- \cA_{\alpha L}^{M \beta} \cA_{\beta K}^{N \alpha}
+ \cA_{\alpha L}^{N \beta} \cA_{\beta K}^{M \alpha}
\eea 
relating one-loop graphs on the left-hand side to tree-level graphs on the right-hand side.

\sm

For genus $h=2$, the anti-symmetry separately in $MN$ and in $KL$ of (\ref{weight2}) allows us to uniquely contract with the combination $\ep^{KL} \ep_{MN}$, so that the tensorial identity is actually an invariant. It will be shown in section \ref{sec:5.1} that this identity is identical to the one proven for genus two in appendix D of \cite{DHoker:2020tcq}. For arbitrary genus, there is a single inequivalent contraction of indices in (\ref{weight2}), since both sides are anti-symmetric separately in $MN$ and in $KL$. Contracting $M$ with $K$ gives,
\bea
 \cB_{\alpha L} ^{\alpha N} - \cB_{\alpha L} ^{ N \alpha } 
& = & 
\cA^{N \b \a}_{\a \b L} + \cA^{\b N \a}_{\a L \b} - \cA^{ \b N \a }_{\a \b L}  - \cA^{N \b \a }_{\a L \b}
- \cA^{\gamma \beta}_{\a L} \cA^{ N \a }_{\b \gamma} + \cA^{ N \b }_{ \a L }  \cA^{\gamma \alpha}_{\beta \gamma} 
\label{weight2c}
\eea
Contracting also $L$ with $N$ gives,
\bea
 \cB_{\alpha \beta} ^{\alpha \beta} - \cB_{\alpha \beta} ^{\beta \alpha} 
& = & 
2\cA^{\gamma \b \a}_{\a \b \gamma} - 2\cA^{\b \gamma \a}_{\a \b \gamma} 
- \cA_{\alpha \delta } ^{ \gamma \b } \cA_{\b \gamma}^{\delta \a} + \cA_{\a \delta } ^{\delta \beta}  \cA_{\b \gamma} ^{\gamma\a} 
\label{weight2cc}
\eea
As will be detailed in section \ref{sec:5.1}, this identity generalizes the genus-two identity of appendix D in \cite{DHoker:2020tcq} to arbitrary genus.

\subsection{Weight 3}

For weight 3, namely $n=1$, the tensor in Theorem \ref{thm0} reduces to,
\begin{align}
\cT^{JNM}_{\, ILK}&=\frac{i}{2} \int_{\Sigma^2_{x,y}} W_{IL}^J(x,y) \Big ( \cG(x,y) \mu_K^M(x) 
- \mu _\a ^M(x) \Phi _K^\a(y) \Big ) \oom^N(y)\notag
\\
& = \int_{\Sigma^2_{x,y}}  
\Big ( \cG(x,y) \mu_K^M(x) 
- \mu _\a ^M(x) \Phi _K^\a(y) \Big ) \label{ids.2} 
\\
&\ \ \ \ \ \ \times \Big(
 V_I^J(x,y) \mu_L^N(y) - \Phi_I^\beta(x) \Phi_\beta^J(y)  \mu_L^N(y) 
-   \Phi _{IL}^{J\beta} (x)  \mu_\beta^N(y) + \Phi_I^\beta(x) \cA_{\beta L}^{J\gamma} \mu^N_\gamma(y)
\Big) \notag 
\end{align}
where we have used the expression (\ref{otherW2}) for $W_{IL}^J(x,y)$. We again perform the integrals using the recursive definitions of various tensors and rename the indices $(M,J,N) \rightarrow (J_1,J_2,J_3)$ and $(K,I,L) \rightarrow (I_1,I_2,I_3)$ in order to make the cyclicity  $\cT^{J_1 J_2 J_3}_{\, I_1 I_2 I_3}  = \cT^{J_3 J_1 J_2 }_{ \,I_3 I_1 I_2 } $ of the following simplified expression more transparent,
\begin{align}
\cT^{J_1 J_2 J_3}_{\, I_1 I_2 I_3} &= \cB^{J_1 J_2 J_3}_{I_1 I_2 I_3} - \cA^{J_1 J_2 J_3 \alpha}_{\alpha I_2 I_3 I_1}
- \cA^{J_2 J_3 J_1 \alpha}_{\alpha I_3 I_1 I_2} - \cA^{J_3 J_1 J_2 \alpha}_{\alpha I_1 I_2 I_3} \notag \\
&
+ \cA^{J_3 \beta}_{\alpha I_1}  \cA^{J_1 J_2 \alpha}_{\beta I_2 I_3 }
+ \cA^{J_1 \beta}_{\alpha I_2}  \cA^{J_2 J_3 \alpha}_{\beta I_3 I_1 }
+ \cA^{J_2 \beta}_{\alpha I_3}  \cA^{J_3 J_1 \alpha}_{\beta I_1 I_2 }
- \cA^{J_3 \beta}_{\alpha I_1}  \cA^{J_1 \gamma}_{\beta I_2}  \cA^{J_2 \alpha}_{\gamma I_3 }
\label{ids.3} 
\end{align}
While there is no obvious analogue of the reflection identity
$\cB^{J_1 J_2 J_3}_{I_1 I_2 I_3} 
= \cB^{J_3 J_2 J_1}_{I_3 I_2 I_1} $ for the tensor $\cT$, Theorem \ref{thm0} implies (\ref{ids.2}) to be 
symmetric under simultaneous exchange $M\leftrightarrow N$ and $I \leftrightarrow L$. 
This amounts to the following symmetry property of the tensor in (\ref{ids.3}),
\bea
\cT^{J_1 J_2 J_3}_{\,I_1 I_2 I_3} = \cT^{J_2 J_1 J_3}_{\,I_3 I_2 I_1}
\label{weight3}
\eea
on top of its cyclicity. As in the weight-two case (\ref{weight2}), this identity relates a combination of one-loop
modular graph tensors $\cB^{J_1 J_2 J_3}_{I_1 I_2 I_3}-\cB^{J_2 J_1 J_3}_{I_3 I_2 I_1}$
to tree-level ones, see section \ref{sec:5.2} for the corollaries for modular graph functions at weight three.

\subsection{Weight 4}

For weight 4, namely $n=2$, the tensor in Theorem \ref{thm0} reduces to a relabelling of
\begin{align}
\cT^{J_1 J_2 J_3 J_4}_{\, I_1 I_2 I_3 I_4} &= \cB^{J_1 J_2 J_3 J_4}_{I_1 I_2 I_3 I_4}
 - \cA^{J_1 J_2 J_3 J_4 \alpha}_{\alpha I_2 I_3 I_4 I_1}
  - \cA^{J_2 J_3 J_4 J_1 \alpha}_{\alpha I_3 I_4 I_1 I_2}
   - \cA^{J_3 J_4 J_1 J_2 \alpha}_{\alpha I_4 I_1 I_2 I_3}
    - \cA^{J_4 J_1 J_2 J_3 \alpha}_{\alpha I_1 I_2 I_3 I_4}
  \notag \\
&
+ \cA^{J_4 \beta}_{\alpha I_1}  \cA^{J_1 J_2 J_3 \alpha}_{\beta I_2 I_3 I_4}
+ \cA^{J_1 \beta}_{\alpha I_2}  \cA^{J_2 J_3 J_4 \alpha}_{\beta I_3 I_4 I_1}
+ \cA^{J_2 \beta}_{\alpha I_3}  \cA^{J_3 J_4 J_1 \alpha}_{\beta I_4 I_1 I_2}
+ \cA^{J_3 \beta}_{\alpha I_4}  \cA^{J_4 J_1 J_2 \alpha}_{\beta I_1 I_2 I_3}
\notag \\
&
+ \cA^{J_3 J_4 \beta}_{\alpha I_4 I_1}  \cA^{J_1 J_2 \alpha}_{\beta I_2 I_3}
+ \cA^{J_4 J_1 \beta}_{\alpha I_1 I_2}  \cA^{J_2 J_3 \alpha}_{\beta I_3 I_4}
+ \cA^{J_1 \beta}_{\alpha I_2} \cA^{J_2 \gamma}_{\beta I_3}\cA^{J_3 \delta}_{\gamma I_4}\cA^{J_4 \alpha}_{\delta I_1}
\label{ids.7}  \\
&
- \cA^{J_1 \beta}_{\alpha I_2} \cA^{J_2 \gamma}_{\beta I_3}\cA^{J_3 J_4 \alpha}_{\gamma I_4 I_1}
- \cA^{J_2 \beta}_{\alpha I_3} \cA^{J_3 \gamma}_{\beta I_4}\cA^{J_4 J_1 \alpha}_{\gamma I_1 I_2}
- \cA^{J_3 \beta}_{\alpha I_4} \cA^{J_4 \gamma}_{\beta I_1}\cA^{J_1 J_2 \alpha}_{\gamma I_2 I_3}
- \cA^{J_4 \beta}_{\alpha I_1} \cA^{J_1 \gamma}_{\beta I_2}\cA^{J_2 J_3 \alpha}_{\gamma I_3 I_4}
\notag
\end{align}
which again enjoys cyclicity $\cT^{J_1 J_2 J_3 J_4}_{\, I_1 I_2 I_3 I_4} = \cT^{J_4 J_1 J_2 J_3}_{\,I_4 I_1 I_2 I_3}$ but no obvious reflection property. The 16 terms in (\ref{ids.7}) are
assembled from the contributions (\ref{CDS.3}) to $W^{J_1 J_2}_{I_1 I_2 L}(x,y)$. As a result of Theorem \ref{thm0},
the tensor obeys the symmetry
\bea
\cT^{J_1 J_2 J_3 J_4}_{\, I_1 I_2 I_3 I_4} = \cT^{J_3 J_2 J_1 J_4}_{\,I_4 I_3 I_2 I_1}
\label{ids.8} 
\eea
and once more relates a difference of one-loop modular graph tensors to tree-level ones.

\subsection{Weight 5}

For weight 5, namely $n=3$, the tensor Theorem \ref{thm0} admits the 
cyclically symmetric representation
\begin{align}
\cT^{J_1 J_2 J_3 J_4 J_5}_{\, I_1 I_2 I_3 I_4 I_5} &= \cB^{J_1 J_2 J_3 J_4 J_5}_{I_1 I_2 I_3 I_4 I_5}
- \cA^{J_1 \beta}_{\alpha I_2} \cA^{J_2 \gamma}_{\beta I_3} \cA^{J_3 \delta}_{\gamma I_4} \cA^{J_4 \epsilon}_{\delta I_5} \cA^{J_5 \alpha}_{\epsilon I_1} \notag \\
&+ \Big\{  {-} \cA^{J_1 J_2 J_3 J_4 J_5 \alpha}_{\alpha I_2 I_3 I_4 I_5 I_1}
+ \cA^{J_1 \beta}_{\alpha I_2}  \cA^{J_2 J_3 J_4 J_5 \alpha}_{\beta I_3 I_4 I_5 I_1}
+ \cA^{J_1 J_2 \beta}_{\alpha I_2 I_3}  \cA^{ J_3 J_4 J_5 \alpha}_{\beta I_4 I_5 I_1} \notag \\
& \ \ \ \ 
- \cA^{J_1 \beta}_{\alpha I_2} \cA^{J_2 \gamma}_{\beta I_3}\cA^{J_3 J_4 J_5 \alpha}_{\gamma I_4 I_5 I_1}
- \cA^{J_1 \beta}_{\alpha I_2} \cA^{J_2 J_3 \gamma}_{\beta I_3 I_4}\cA^{J_4 J_5 \alpha}_{\gamma  I_5 I_1} \notag \\
& \ \ \ \ + \cA^{J_1 \beta}_{\alpha I_2} \cA^{J_2 \gamma}_{\beta I_3}\cA^{J_3  \delta}_{\gamma I_4} \cA^{J_4 J_5 \alpha}_{\delta I_5 I_1} + {\rm cyc}\big(\begin{smallmatrix} J_1, & \! J_2, &\! J_3, &\! J_4, &\! J_5 \\
I_1, &\! I_2, &\! I_3, &\! I_4, &\! I_5 \end{smallmatrix}  \big) \Big\}
\label{ids.9}
\end{align}
which has been obtained on the basis of (\ref{CDS.4}). Theorem \ref{thm0} equates 
(\ref{ids.9}) to the following permutation of its indices:
\bea
\cT^{J_1 J_2 J_3 J_4 J_5}_{ \, I_1 I_2 I_3 I_4 I_5} 
= \cT^{ J_4 J_3 J_2 J_1 J_5}_{\, I_5 I_4 I_3 I_2 I_1}
\label{ids.10}
\eea

\subsection{Weight $n$}
\label{sec3.6}

At general weight $n$, the tensor in Theorem \ref{thm0} can be simplified to the
following cyclically symmetric combination of modular graph tensors
\begin{align}
\cT^{J_1 J_2 \dots J_n}_{\, I_1 I_2 \ldots I_n} &= \cB^{J_1 J_2 \dots J_n}_{I_1 I_2 \ldots I_n} + \sum_{k=1}^n (-1)^k
\sum_{1\leq i_1<i_2<\ldots <i_k \leq n}
\cA^{J_{i_1} J_{i_1+1} \ldots J_{i_2-1} \alpha_2}_{\, \alpha_1 \, I_{i_1+1} \ldots I_{i_2-1} I_{i_2} }
\cA^{J_{i_2} J_{i_2+1} \ldots J_{i_3-1} \alpha_3}_{\, \alpha_2 \, I_{i_2+1} \ldots I_{i_3-1} I_{i_3} } \notag \\
& \ \  \ \  \ \  \ \  \ \  \ \  \ \  \ \ \ \  \ \  \ \  \ \  \ \  \ \  \times \ldots 
\cA^{J_{i_{k-1}} J_{i_{k-1}+1} \ldots J_{i_k-1} \alpha_k}_{\, \alpha_{k-1} \, I_{i_{k-1}+1} \ldots I_{i_k-1} I_{i_k} } 
\cA^{J_{i_k} J_{i_k+1} \ldots J_n J_1  \ldots J_{i_1-1} \alpha_1}_{\, \alpha_k \, I_{i_k+1} \ldots \, I_n I_1\, \ldots I_{i_1-1} I_{i_1} } 
\label{ids.11}
\end{align}
which enjoys the following symmetry by the Theorem
\bea
\cT^{J_1 J_2 \dots J_n}_{\,I_1 I_2 \ldots I_n} = \cT^{ J_{n-1} J_{n-2} \ldots  J_2 J_1 J_n}_{ \  I_n \ \ I_{n-1} \ldots \,  I_3 I_2  I_1} 
\label{ids.12}
\eea
Together with the first term $\cB^{J_1  \dots J_n}_{I_1  \ldots I_n}$ on the right side of (\ref{ids.11}),
the sum $\sum_{k=1}^n \sum_{1\leq i_1<i_2<\ldots <i_k \leq n}$ yields all the $2^n$ possibilities
to replace a subset of the pairs $\begin{smallmatrix} J_k \\ I_k \end{smallmatrix}$ by a contraction
$\cA^{\ldots \alpha}_{\ldots I_k} \cA_{\alpha \ldots }^{J_k \ldots}$. Terms with an odd number of
$\cA$-factors enter with a minus sign. The proof that the left-hand side of (\ref{thm0.1}) leads to (\ref{ids.11})
is based on the representation (\ref{CDS.7}) of the $W$-tensors and given in appendix \ref{app:B}.

\sm

The analogous expressions for the complex conjugate tensors (\ref{cctheorem}) are given by,
\begin{align}
\overline{\cT}^{J_1 J_2 \dots J_n}_{I_1 I_2 \ldots I_n} &= \cB^{J_1 J_2 \dots J_n}_{I_1 I_2 \ldots I_n} + \sum_{k=1}^n (-1)^k
\sum_{1\leq i_1<i_2<\ldots <i_k \leq n}
\cA_{I_{i_1} I_{i_1+1} \ldots I_{i_2-1} \alpha_2}^{ \alpha_1 J_{i_1+1} \ldots J_{i_2-1} J_{i_2} }
\cA_{I_{i_2} I_{i_2+1} \ldots I_{i_3-1} \alpha_3}^{ \alpha_2  J_{i_2+1} \ldots J_{i_3-1} J_{i_3} } \notag \\
& \ \  \ \  \ \  \ \  \ \  \ \  \ \  \ \ \ \  \ \  \ \  \ \  \ \  \ \  \times \ldots 
\cA_{I_{i_{k-1}} I_{i_{k-1}+1} \ldots I_{i_k-1} \alpha_k}^{ \alpha_{k-1}  J_{i_{k-1}+1} \ldots J_{i_k-1} J_{i_k} } 
\cA_{I_{i_k} I_{i_k+1} \ldots \, I_n I_1  \ldots \, I_{i_1-1} \alpha_1}^{ \alpha_k J_{i_k+1} \ldots \, J_n J_1\, \ldots J_{i_1-1} J_{i_1} } 
\label{ccids.11}
\end{align}
whose symmetry properties (\ref{ccthm}) can be rewritten as $\overline{\cT}^{J_1 J_2 \dots J_n}_{I_1 I_2 \ldots I_n} = \overline{\cT}_{ I_{n-1} I_{n-2} \ldots  I_2 I_1 I_n}^{ \ J_n \;  J_{n-1} \ldots   J_3 J_2  J_1} $.

\newpage

\section{Corollaries for scalar modular graph functions}
\label{sec:5}
\setcounter{equation}{0}

In this section, we will extract new identities among the scalar modular graph functions 
in the low energy expansion of higher-genus string amplitudes
by contracting the free indices in the identities (\ref{ids.11}) and (\ref{ids.12}) among modular graph tensors.
Hence, modular graph tensors turn out to be crucial auxiliary objects in the
simplifications of higher-genus string amplitudes in the same way as modular graph forms with
non-trivial modular weights have key input on identities among modular graph 
functions at genus one \cite{DHoker:2016mwo}.

\subsection{Weight 2}
\label{sec:5.1}

For string amplitudes at genus $h=2$, the modular graph functions in their low energy expansions
can be rewritten in terms of the genus-agnostic forms
\bea
\kappa(x) = \frac{i}{2h} \omega_I(x) \overline \omega^I(x) = \frac{1}{h} \mu_I^I(x) \, , \ \ \ \ \ \
\nu(x,y) = \frac{i}{2} \omega_I(x) \overline \omega^I(y)
\label{coro.1}
\eea
without reference to the object $\Delta(x,y) = \omega_1(x) \omega_2(y) - \omega_1(y)\omega_2(x)$ specific to genus two. In particular, the modular graph functions at the subleading low energy  orders \cite{D'Hoker:2013eea, DHoker:2017pvk, DHoker:2018mys, DHoker:2020tcq} admit the following genus-agnostic representations. The Kawazumi-Zhang invariant may be expressed in the following equivalent manners, 
\bea
\varphi = \int_{\Sigma^2} \cG(1,2) \nu(1,2) \nu(2,1) =  \int_{\Sigma^2} \mu_I^J(1)\cG(1,2) \mu_J^I(2) = \cA^{\alpha \beta}_{\beta \alpha} 
\eea
while the remaining invariants are defined as follows, 
\bea
\label{coro.2}
\cZ_1 & = & 8 \int_{\Sigma^2} \cG(1,2)^2 \kappa(1) \kappa(2) 
\no \\
\cZ_2 &= & -4 \int_{\Sigma^3} \cG(1,3) \cG(2,3)  \nu(1,2) \nu(2,1) \kappa(3) 
\no \\ %
\cZ_2' &=& -4 \int_{\Sigma^3} \cG(1,3) \cG(2,3)  \nu(1,2) \nu(2,3) \nu(3,1)
\no \\
\cZ_3 &= & 2 \int_{\Sigma^4} \cG(1,2) \cG(3,4)  \nu(1,3) \nu(3,1)  \nu(2,4) \nu(4,2) 
\no \\
\cZ_3' &= & 2 \int_{\Sigma^4} \cG(1,2) \cG(3,4)  \nu(1,2) \nu(2,3)  \nu(3,4) \nu(4,1) 
\no \\
\cZ_4 &= & - 4 \int_{\Sigma^2} \cG(1,2)^2 \nu(1,2) \nu(2,1) 
\eea
with $\cG(i,j) = \cG(z_i,z_j)$ and the same shorthands for the arguments of $\kappa,\nu$ and $\mu$.
It is straightforward to re-express these integrals in terms of the form $\mu$ in (\ref{mu}), as follows, 
\bea
\label{coro.3}
\cZ_1 &= & \frac{8}{h^2}  \int_{\Sigma^2} \mu_I^I(1) \cG(1,2)^2 \mu_J^J(2)
\no \\
\cZ_2 &= &  - \frac{4}{h}  \int_{\Sigma^3} \mu_I^J(1) \mu_J^I(2) \cG(1,3) \cG(2,3) \mu_K^K(3) 
\no \\
\cZ_2' & = & - 4  \int_{\Sigma^3} \mu_I^K(1) \mu_J^I(2)  \mu_K^J(3) \cG(1,3) \cG(2,3) 
\no \\
\cZ_3 &= & 2 \int_{\Sigma^4} \cG(1,2) \cG(3,4) \mu_I^J(1) \mu_J^I(3) \mu_K^L(2) \mu_L^K(4)
\no \\
 \cZ_3' &= &  2 \int_{\Sigma^4} \cG(1,2) \cG(3,4) \mu_I^L(1) \mu^I_J(2) \mu^J_K(3) \mu_L^K(4)
\no \\
\cZ_4 &= &  -4   \int_{\Sigma^2} \mu_I^J(1)\cG(1,2)^2 \mu_J^I(2) 
\eea
The modular graph functions in (\ref{coro.2}) and (\ref{coro.3}) can be identified
as contractions of the tensors $\cA$ and $\cB$ associated with tree-level and
one-loop graphs, respectively,
\begin{align}
%
\cZ_2 &= - \frac{4}{h} \cA^{\alpha \beta \gamma}_{\gamma \beta \alpha} 
&\cZ_2' &= - 4 \cA^{\alpha \beta \gamma}_{\gamma \alpha \beta} 
\label{coro.4} \\
\cZ_3 &= 2 \cA^{\alpha \beta}_{\gamma \delta} \cA^{\delta \gamma}_{\beta \alpha}
&\cZ_3' &= 
2 \cA^{\gamma \beta}_{\alpha \gamma} \cA^{\delta \alpha}_{\beta \delta}
\notag \\
\cZ_1 &= \frac{8}{h^2} \cB^{\alpha \beta}_{\alpha \beta} 
&\cZ_4 &= -4 \, \cB^{\alpha \beta}_{\beta \alpha} \notag
\end{align}
These modular graph tensors at weight two have already
been related by contracting the indices in the identity (\ref{weight2}).
As a consequence of (\ref{weight2cc}), the genus-$h$ modular
graph functions in (\ref{coro.4}) are related by
\bea
\frac{h^2}{4} \cZ_1 + h \cZ_2 - \cZ_2' + \cZ_3 - \cZ_3' + \frac{1}{2} \cZ_4 = 0 
\label{coro.5}
\eea
This generalizes the $h=2$ relation identified and proven in \cite{DHoker:2020tcq}
to arbitrary genus. In order to recover the genus-two identity in the reference,
we exploit relations such $\cA^{[\alpha \beta \gamma]}_{\phantom{[} \alpha \beta \gamma \phantom{]}}=0$ 
as specific to $h=2$ (due to the vanishing of anti-symmetrizations in $h+1$ indices),
\bea
 h=2 \ \ \Rightarrow \ \ \left\{ \begin{array}{ll}
0 = \cA^{[\alpha \beta \gamma]}_{\phantom{[} \alpha \beta \gamma \phantom{]}}
&\Rightarrow \ \ \cZ_2' = \cZ_2 \\
0 =  \cA^{\alpha [ \beta}_{\beta \phantom{[} \alpha} \cA^{\gamma \delta]}_{\gamma \delta}
&\Rightarrow \ \ \cZ_3' =\varphi^2
\end{array} \right. 
\label{coro.6}
\eea
This yields the special case
\bea
h=2 \ \ \Rightarrow \ \  \cZ_1 +  \cZ_2 + \cZ_3  + \frac{1}{2} \cZ_4 - \varphi^2 =0
\label{coro.7}
\eea
of (\ref{coro.5}) which was derived and applied to the simplification of genus two five-point
superstring amplitudes in \cite{DHoker:2020tcq}. We emphasize that (\ref{coro.6}) and
(\ref{coro.7}) no longer hold at $h\geq 3$.

\subsection{Weight 3}
\label{sec:5.2}

For the weight-three identity (\ref{weight3}) among the tensors $\cT^{J_1 J_2 J_3}_{I_1 I_2 I_3}$
in (\ref{ids.3}), there is again a unique way of deriving a non-trivial identity from contracting all indices.
This leads to the eight-term identity,%
\bea
h^3 \cD_1 - \cD_2 - 3 h^2 \cD_3 + 3 \cD_4 + 3h \cD_5 - 3 \cD_6 - \cD_7 + \cD_8 =0
\label{coro.8}
\eea
among the following modular graph functions involving up to six integrated punctures
\begin{align}
\cD_1 &= \frac{1}{h^3} \cB^{\alpha \beta \gamma}_{\alpha \beta \gamma} = \int_{\Sigma^3} \cG(1,2) \cG(2,3) \cG(3,1) \kappa(1) \kappa(2) \kappa(3)
\notag \\
\cD_2 &=  \cB^{\alpha \beta \gamma}_{\gamma \alpha \beta} = \int_{\Sigma^3} \cG(1,2) \cG(2,3) \cG(3,1) 
\nu(1,2)\nu(2,3)\nu(3,1)
\notag \\
\cD_3 &= \frac{1}{h^2} \cA^{\beta \gamma \delta \alpha}_{ \alpha \gamma \delta \beta} = \int_{\Sigma^4} \cG(1,2) \cG(2,3) \cG(3,4) 
\nu(1,4)\nu(4,1) \kappa(2) \kappa(3)
\notag \\
\cD_4 &=   \cA_{\beta \gamma \delta \alpha}^{ \alpha \beta \gamma \delta} = \int_{\Sigma^4} \cG(1,2) \cG(2,3) \cG(3,4) 
\nu(1,2)\nu(2,3) \nu(3,4) \nu(4,1) 
\label{coro.11} \\
\cD_5 &=  \frac{1}{h}  \cA^{\gamma \beta}_{ \alpha \delta} \cA^{\delta \epsilon \alpha}_{\beta \epsilon \gamma} = \int_{\Sigma^5} \cG(1,2) \cG(3,4) \cG(4,5) 
\nu(2,3)\nu(3,2) \nu(1,5) \nu(5,1) \kappa(4) 
\notag \\
\cD_6 &=   \cA^{\gamma \beta}_{ \alpha \gamma} \cA^{\delta \epsilon \alpha}_{\beta \delta \epsilon} = \int_{\Sigma^5} \cG(1,2) \cG(3,4) \cG(4,5) 
\nu(1,2)\nu(2,3) \nu(3,4) \nu(4,5) \nu(5,1)
\notag \\
\cD_7 &=   \cA^{\pi \beta}_{ \alpha \delta}  \cA^{\delta \gamma}_{ \beta \epsilon}  \cA^{\epsilon \alpha }_{ \gamma \pi} 
= \int_{\Sigma^6} \cG(1,2) \cG(3,4) \cG(5,6) 
\nu(2,3)\nu(3,2) \nu(4,5) \nu(5,4) \nu(6,1) \nu(1,6)
\notag \\
\cD_8 &=   \cA^{\delta \beta}_{ \alpha \delta}  \cA^{\epsilon \gamma}_{ \beta \epsilon}  \cA^{\pi \alpha }_{ \gamma \pi} 
= \int_{\Sigma^6} \cG(1,2) \cG(3,4) \cG(5,6) 
\nu(1,2)\nu(2,3) \nu(3,4) \nu(4,5) \nu(5,6) \nu(6,1)
\notag
\end{align}
The $h=2$ instances of the first six modular graph functions $\cD_1,\cD_2,\ldots,\cD_6$
enter the $D^{10}\cR^4$ and $D^8 \cR^5$ effective interactions of the four-
and five-point genus two superstring amplitudes, and $\cD_7, \cD_8$
are expected to appear at six points at the order of $D^6 \cR^6$.
When extending the graphical bookkeeping of section \ref{sec:graph} to $\kappa(i)$ and $\nu(i,j)$,
%
\[
\tikzpicture[scale=1]
\scope[xshift=0cm,yshift=0cm]
\draw(-1.5,0)node{$\cG(x,y) =$};
\draw (0,0)node[left]{$x$} -- (1,0)node[right]{$y$};
\draw[fill=white] (0,0)  circle [radius=.07] ;
\draw[fill=white] (1,0)  circle [radius=.07] ;
\draw(1.6,0)node{$,$};
\draw(3.5,0)node{$\nu(x,y) =$};
\draw[dashed] (5,0)node[left]{$x$} -- (6,0)node[right]{$y$};
\draw[->](5.5,0)--(5.51,0);
\draw[fill=white] (5,0)  circle [radius=.07] ;
\draw[fill=white] (6,0)  circle [radius=.07] ;
\draw(6.6,0)node{$,$};
\draw(8.5,0)node{$\kappa(x) =$};
\draw[dashed](9.5,0)node[above]{$x$} ..controls (10.5,1) and (10.5,-1) .. (9.5,0) ;
\draw[->](10.24,0)--(10.24,0.01);
\draw[fill=white] (9.5,0)  circle [radius=.07] ;
\endscope
\endtikzpicture
\]
%
the modular graph functions $\cD_{j}$ in (\ref{coro.11}) can be visualized as follows:

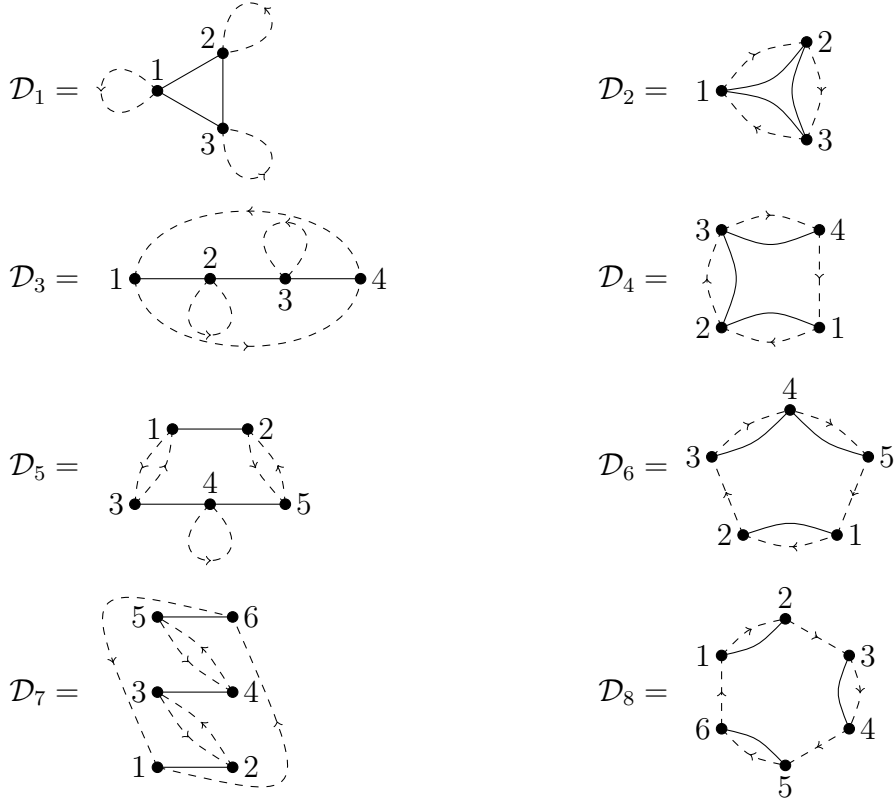
\begin{center}
\tikzpicture[scale=1]
\scope[xshift=0cm,yshift=0cm]
\draw(-1.5,0)node{$\cD_1=$};
\draw(0,0) -- (0.87,0.5);
\draw(0,0) -- (0.87,-0.5);
\draw(0.87,0.5) -- (0.87,-0.5);
\draw[dashed](0,0)node[above]{$1$}   ..controls (-1,1) and (-1,-1) .. (0,0) ;
\draw[->](-0.75,0)--(-0.75,-0.01);
\draw[dashed](0.87,0.5)   ..controls (0.87,1.9) and (2.34,0.7) .. (0.87,0.5) ;
\draw[->](1.425,1.105)--(1.424,1.106);
\draw[dashed](0.87,-0.5)   ..controls (0.87,-1.9) and (2.34,-0.7) .. (0.87,-0.5) ;
\draw[->](1.425,-1.105)--(1.426,-1.104);
\draw(0.67,0.7)node{$2$};
\draw(0.67,-0.7)node{$3$};
\draw[fill=black] (0,0)  circle [radius=.07] ;
\draw[fill=black] (0.87,0.5)  circle [radius=.07] ;
\draw[fill=black] (0.87,-0.5)  circle [radius=.07] ;
\endscope
\scope[xshift=7.5cm,yshift=0cm,scale=1.3]
\draw(-0.9,0)node{$\cD_2=$};
\draw[dashed](0,0) node[left]{$1$} ..controls (0.435-0.1,0.25+0.174) .. (0.87,0.5) node[right]{$2$};
\draw(0,0) ..controls (0.435+0.1,0.25-0.174) .. (0.87,0.5);
\draw[->] (0.338,0.374) -- (0.34,0.375);
\draw[dashed](0,0) ..controls (0.435-0.1,-0.25-0.174) .. (0.87,-0.5) node[right]{$3$};
\draw(0,0) ..controls (0.435+0.1,-0.25+0.174) .. (0.87,-0.5);
\draw[->] (0.342,-0.376) -- (0.34,-0.375);
\draw(0.87,0.5) ..controls (0.67,0) .. (0.87,-0.5);
\draw[dashed](0.87,0.5) ..controls (1.07,0) .. (0.87,-0.5);
\draw[->] (1.023,0)--(1.023,-0.01);
\draw[fill=black] (0,0)  circle [radius=.07/1.3] ;
\draw[fill=black] (0.87,0.5)  circle [radius=.07/1.3] ;
\draw[fill=black] (0.87,-0.5)  circle [radius=.07/1.3] ;
\endscope
\scope[xshift=0cm,yshift=-2.5cm]
\draw(-1.5,0)node{$\cD_3=$};
\scope[xshift=-0.3cm]
\draw(0,0)node[left]{$1$} -- (3,0)node[right]{$4$};
\draw[dashed](1,0)node[above]{$2$}   ..controls (0,-1) and (2,-1) .. (1,0) ;
\draw[dashed](2,0)node[below]{$3$}   ..controls (1,1) and (3,1) .. (2,0) ;
\draw[->](1,-0.75)--(1.01,-0.75);
\draw[->](2.01,0.75)--(2,0.75);
\draw[dashed](0,0)  ..controls (0.1,1.2) and (2.9,1.2) .. (3,0) ;
\draw[dashed](0,0)  ..controls (0.1,-1.2) and (2.9,-1.2) .. (3,0) ;
\draw[->](1.51,0.9)--(1.5,0.9);
\draw[->](1.5,-0.9)--(1.51,-0.9);
\draw[fill=black] (0,0)  circle [radius=.07];
\draw[fill=black] (1,0)  circle [radius=.07] ;
\draw[fill=black] (2,0)  circle [radius=.07] ;
\draw[fill=black] (3,0)  circle [radius=.07];
\endscope
\endscope
\scope[xshift=7.5cm,yshift=-2.5cm, scale=1.3]
\draw(-0.9,0)node{$\cD_4=$};
\draw[dashed](0,-0.5)  ..controls (-0.2,0) .. (0,0.5) ;
\draw(0,-0.5)  ..controls (0.2,0) .. (0,0.5) ;
\draw[<-](-0.15,0)--(-0.15,-0.01);
\draw[dashed](0,0.5)  ..controls (0.5,0.7) .. (1,0.5) ;
\draw(0,0.5)  ..controls (0.5,0.3) .. (1,0.5) ;
\draw[<-](0.5,0.65)--(0.49,0.65);
\draw[dashed](0,-0.5)  ..controls (0.5,-0.7) .. (1,-0.5) ;
\draw(0,-0.5)  ..controls (0.5,-0.3) .. (1,-0.5) ;
\draw[<-](0.5,-0.65)--(0.51,-0.65);
\draw[dashed](1,0.5) -- (1,-0.5) ;
\draw[<-](1,0)--(1,0.01);
\draw[fill=black] (0,-0.5)  circle [radius=.07/1.3]  node[left]{$2$};
\draw[fill=black] (0,0.5)  circle [radius=.07/1.3] node[left]{$3$};
\draw[fill=black] (1,0.5)  circle [radius=.07/1.3] node[right]{$4$};
\draw[fill=black] (1,-0.5)  circle [radius=.07/1.3] node[right]{$1$};
\endscope
\scope[xshift=0cm,yshift=-5cm]
\draw(-1.5,0)node{$\cD_5=$};
\scope[xshift=-0.3cm]
\draw(0.5,0.5)node[left]{$1$} -- (1.5,0.5)node[right]{$2$};
\draw(0,-0.5)node[left]{$3$} -- (2,-0.5)node[right]{$5$};
\draw[dashed](1,-0.5)node[above]{$4$}   ..controls (0,-1.5) and (2,-1.5) .. (1,-0.5) ;
\draw[->](1,-1.25)--(1.01,-1.25);
\draw[dashed](0.5,0.5)   ..controls (0.05,0) .. (0,-0.5) ;
\draw[dashed](0.5,0.5)   ..controls (0.45,0) .. (0,-0.5) ;
\draw[->](0.11,0)--(0.1,-0.02);
\draw[->](0.395,0)--(0.405,0.02);
\draw[dashed](1.5,0.5)   ..controls (1.95,0) .. (2,-0.5) ;
\draw[dashed](1.5,0.5)   ..controls (1.55,0) .. (2,-0.5) ;
\draw[->](1.91,0)--(1.9,0.02);
\draw[->](1.595,0)--(1.605,-0.02);
\draw[fill=black] (0.5,0.5)  circle [radius=.07];
\draw[fill=black] (1.5,0.5)  circle [radius=.07] ;
\draw[fill=black] (0,-0.5)  circle [radius=.07] ;
\draw[fill=black] (1,-0.5)  circle [radius=.07] ;
\draw[fill=black] (2,-0.5)  circle [radius=.07] ;
\endscope
\endscope
\scope[xshift=7.5cm,yshift=-5cm, scale=1.3]
\draw(-0.9,0)node{$\cD_6=$};
\scope[xshift=0.7cm,yshift=0.1cm,scale=0.8]
\draw[dashed](-0.6,-1)  ..controls (0,-1.2) .. (0.6,-1) ;
\draw(-0.6,-1)  ..controls (0,-0.8) .. (0.6,-1) ;
\draw[<-](0.03,-1.15)--(0.04,-1.15);
\draw[dashed](0.6,-1)  -- (1,0) ;
\draw[dashed](-0.6,-1)  -- (-1,0) ;
\draw[<-](0.8,-0.5)--(0.82,-0.45);
\draw[<-](-0.82,-0.45)--(-0.8,-0.5);
\draw[dashed](1,0).. controls (0.5+0.06,0.3+0.14).. (0,0.6);
\draw(1,0).. controls (0.5-0.06,0.3-0.14).. (0,0.6);
\draw[<-](0.54,0.41)--(0.52,0.42);
\draw[dashed](-1,0).. controls (-0.5-0.06,0.3+0.14).. (0,0.6);
\draw(-1,0).. controls (-0.5+0.06,0.3-0.14).. (0,0.6);
\draw[<-](-0.52,0.42)--(-0.54,0.41);
\draw[fill=black] (0,0.6)  circle [radius=.07/1.3/0.8]  node[above]{$4$};
\draw[fill=black] (1,0)  circle [radius=.07/1.3/0.8] node[right]{$5$};
\draw[fill=black] (-1,0)  circle [radius=.07/1.3/0.8] node[left]{$3$};
\draw[fill=black] (-0.6,-1)  circle [radius=.07/1.3/0.8] node[left]{$2$};
\draw[fill=black] (0.6,-1)  circle [radius=.07/1.3/0.8] node[right]{$1$};
\endscope
\endscope
\scope[xshift=0cm,yshift=-8cm]
\draw(-1.5,0)node{$\cD_7=$};
\draw(0,-1)node[left]{$1$} -- (1,-1)node[right]{$2$};
\draw(0,0)node[left]{$3$} -- (1,0)node[right]{$4$};
\draw(0,1)node[left]{$5$} -- (1,1)node[right]{$6$};
\draw[dashed](1,-1)   ..controls (0.5+0.12,-0.5+0.12) .. (0,0) ;
\draw[dashed](1,-1)   ..controls (0.5-0.12,-0.5-0.12) .. (0,0) ;
\draw[->](0.4,0.41)--(0.41,0.4);
\draw[->](0.6,0.59)--(0.59,0.6);
\draw[dashed](1,0)   ..controls (0.5+0.12,0.5+0.12) .. (0,1) ;
\draw[dashed](1,0)   ..controls (0.5-0.12,0.5-0.12) .. (0,1) ;
\draw[->](0.4,0.41-1)--(0.41,0.4-1);
\draw[->](0.6,0.59-1)--(0.59,0.6-1);
\draw[dashed](0,-1) ..controls (2.1,-1.5)  .. (1,1);
\draw[dashed](0,-1) ..controls (-1.1,1.5)  .. (1,1);
\draw[->](-0.595,0.41)--(-0.59,0.39);
\draw[->](1+0.595,-0.41)--(1+0.59,-0.39);
\draw[fill=black] (0,0)  circle [radius=.07];
\draw[fill=black] (1,0)  circle [radius=.07];
\draw[fill=black] (0,1)  circle [radius=.07];
\draw[fill=black] (1,1)  circle [radius=.07];
\draw[fill=black] (0,-1)  circle [radius=.07];
\draw[fill=black] (1,-1)  circle [radius=.07];
\endscope
\scope[xshift=7.5cm,yshift=-8cm, scale=1.3]
\draw(-0.9,0)node{$\cD_8=$};
\scope[scale=0.75]
\draw[dashed](0,0.5).. controls (0.43-0.06,0.75+0.14).. (0.87,1);
\draw(0,0.5).. controls (0.43+0.06,0.75-0.14).. (0.87,1);
\draw[->](0.39,0.85)--(0.41,0.86);
\draw[dashed](0,-0.5).. controls (0.43-0.06,-0.75-0.14).. (0.87,-1);
\draw(0,-0.5).. controls (0.43+0.06,-0.75+0.14).. (0.87,-1);
\draw[<-](0.39,-0.85)--(0.41,-0.86);
\draw[dashed](1.74,0.5)-- (0.87,1);
\draw[->](1.3,0.75)--(1.31,0.745);
\draw[dashed](1.74,-0.5)-- (0.87,-1);
\draw[<-](1.3,-0.75)--(1.31,-0.745);
\draw[dashed](0,0.5)--(0,-0.5);
\draw[->](0,0)--(0,0.01);
\draw(1.74,0.5) .. controls (1.54,0) .. (1.74,-0.5);
\draw[dashed](1.74,0.5) .. controls (1.94,0) .. (1.74,-0.5);
\draw[->](1.89,0)--(1.89,-0.01);
\draw[fill=black] (0,0.5)  circle [radius=.07/1.3/0.75]  node[left]{$1$};
\draw[fill=black] (0.87,1)  circle [radius=.07/1.3/0.75] node[above]{$2$};
\draw[fill=black] (1.74,0.5)  circle [radius=.07/1.3/0.75] node[right]{$3$};
\draw[fill=black] (0,-0.5)  circle [radius=.07/1.3/0.75]  node[left]{$6$};
\draw[fill=black] (0.87,-1)  circle [radius=.07/1.3/0.75] node[below]{$5$};
\draw[fill=black] (1.74,-0.5)  circle [radius=.07/1.3/0.75] node[right]{$4$};
\endscope
\endscope
\endtikzpicture
\end{center}

One may again wonder about applications and additional simplifications at low genus. Given
the appearance of the antisymmetric combination $\Delta(x,y,z)= \varepsilon^{IJK}
\omega_I(x) \omega_J(y) \omega_K(z)$ in the genus-three four-point amplitude
\cite{Gomez:2013sla}, it is instructive to rewrite the integral over
$|\Delta(1,2,3)|^2$ in terms of the ${\cal D}_i$ in (\ref{coro.11})
\begin{align}
&\Big( \frac{i}{2}\Big)^3 \int_{\Sigma^3} \cG(1,2) \cG(2,3) \cG(3,1) 
 \varepsilon^{IJK} \omega_I(1) \omega_J(2) \omega_K(3)
 \varepsilon_{PQR} \overline \omega^P(1) \overline \omega^Q(2) \overline \omega^R(3) 
\notag \\
 &\ \ = h^3 \cD_1 + 2 \cD_2 -3 h \int_{\Sigma^3}  \cG(1,2) \cG(2,3) \cG(3,1) \nu(1,2) \nu(2,1) \kappa(3)
  \label{coro.12} 
\end{align}
Hence, the genus-agnostic representation of the $h=3$ integral over $|\Delta(1,2,3)|^2$
introduces another modular graph function with integrand $\sim  \nu(1,2) \nu(2,1) \kappa(3)$
that does not enter the weight-three identity (\ref{coro.8}). At genus two, in turn, the right-hand
side of (\ref{coro.12}) vanishes and relates the integral over $\nu(1,2) \nu(2,1) \kappa(3)$
to $\cD_1$ and $\cD_2$.

By extending the above reasoning to higher weight, the traces of the
tensor identities (\ref{ids.8}), (\ref{ids.10}) or (\ref{ids.12}) generate 
an infinite family of higher-weight generalizations of (\ref{coro.5})
and (\ref{coro.8}) relating weight-$n$ modular graph functions
with up to $2n$ integrated punctures.

\newpage

\section{Further applications of the interchange lemma}
\label{sec:4}
\setcounter{equation}{0}

In this section, we begin to explore further properties of the $W$-tensors and consequences
of the generalized interchange lemma (\ref{lemma2a}). The so-called triangle and square moves to be
spelled out below for conversions of derivatives in a Feynman graph are expected to play a key role
in future work to derive more general identities between modular graph tensors beyond those 
in section \ref{sec:3}.

\subsection{An alternative theorem}
\label{sec:4.1}

This subsection is dedicated to another theorem which leads to an alternative way of deriving the
identities of the previous section.

{\lem
\label{lemma3}
On a compact Riemann surface of  genus $h \geq 1$, the  tensors $W_{I_1 \cdots I_n L}^{J_1 \cdots J_n}(x,y)$ satisfy the following mixed derivative equation which generalizes (\ref{Ara2}),
\bea
\label{lemma3a}
\p_x \pby W_{I_1 \cdots I_n L}^{J_1 \cdots J_n}(x,y) & = & 
\pi  W_{I_n  I_{n-1}\cdots I_1 }^{J_{n-1}  \, \cdots  \, J_1 } (y,x) \, \om_L(y) \, \oom^{J_n}(y)
\no \\ &&
- \pi  \om_L(y) \, \oom^\beta(y) \int _{\Sigma_z} \mu_\beta^{J_n} (z) W_{I_n I_{n-1} \cdots I_1 }^{J_{n-1} \cdots J_1} (z,x) 
\eea
}

\vspace{-0.5cm}
\sm
The proof proceeds by converting the $x$-derivative into a $y$-derivative using Lemma \ref{lemma2} and then 
using the recursion relation that defines $W$ to work out the mixed derivatives. 

{\lem
\label{lemma7}
On a compact Riemann surface of  genus $h \geq 1$, the  tensors $W_{I_1 \cdots I_n L}^{J_1 \cdots J_n}(x,y)$ integrate to zero under the contracted measure $\mu^M_\beta(x) \oom^\beta(y)$,
\bea
\label{lemma7a}
\int _{\Sigma^2}  \mu^M_\beta(x) \,  W_{I_1 \cdots I_n L}^{J_1 \cdots J_n}(x,y) \,\oom^\beta(y) = 0
\eea
}

\vspace{-0.5cm}
\sm
This can be proven based on the representation (\ref{CDS.7}) of the $W$-tensors,
which brings the left-hand side of (\ref{lemma7a}) into the form
\begin{align}
&\int _{\Sigma^2}  \mu^M_\beta(x) \bigg\{
V^{J_1 J_2 \dots J_n}_{\, I_1 I_2\, \ldots \, I_n}(x,y)\mu^\beta_L(y)
- \Phi^{J_1 J_2 \dots J_n K}_{ \, I_1 I_2\,  \ldots \, I_n L}(x) \mu^\beta_K(y)  
\no \\
&\ \ 
+ \sum_{k=1}^n (-1)^k \!\!\!
\sum_{1\leq i_1<i_2<\ldots <i_k \leq n} \!\!\!
\Phi^{J_1  \ldots J_{i_1-1} \alpha_1}_{ I_1\, \ldots I_{i_1-1} I_{i_1}}(x) \, 
\cA^{J_{i_1} J_{i_1+1} \ldots J_{i_2-1} \alpha_2}_{\, \alpha_1 \, I_{i_1+1} \ldots I_{i_2-1} I_{i_2} }
\no \\ & \qquad  \qquad \times 
\cA^{J_{i_2} J_{i_2+1} \ldots J_{i_3-1} \alpha_3}_{\, \alpha_2 \, I_{i_2+1} \ldots I_{i_3-1} I_{i_3} }  
~ \times \cdots \times ~
\cA^{J_{i_{k-1}} J_{i_{k-1}+1} \ldots J_{i_k-1} \alpha_k}_{\, \alpha_{k-1} \, I_{i_{k-1}+1} \ldots I_{i_k-1} I_{i_k} } 
\no \\ & \qquad \qquad \times 
\Big (  
\Phi^{J_n  J_{n-1} \ldots  J_{i_k+1} J_{i_k} }_{ I_n \, I_{n-1} \ldots \,  I_{i_k+1}  \, \alpha_k \,} (y) \mu^\beta_L(y) 
- \cA^{J_{i_k} J_{i_k+1} \ldots J_n K}_{\, \alpha_k \, I_{i_k+1} \ldots \, I_n  L}\mu^\beta_K(y) \Big ) \bigg\}
\end{align}
The first line integrates to $\cA^{M J_1\ldots J_n \beta}_{\beta I_1 \ldots I_n L} -
\cA^{M J_1\ldots J_n K}_{\beta I_1 \ldots I_n L} \delta^\beta_K = 0$ by itself, and the $y$-integral 
over each summand w.r.t.\ $k$ and $i_1,\ldots,i_k$ vanishes separately,
\bea
\int_{\Sigma} \Big(\Phi^{J_n  J_{n-1} \ldots  J_{i_k+1} J_{i_k} }_{ I_n \, I_{n-1} \ldots \,  I_{i_k+1}  \, \alpha_k \,} (y) \mu^\beta_L(y) 
- \cA^{J_{i_k} J_{i_k+1} \ldots J_n K}_{\, \alpha_k \, I_{i_k+1} \ldots \, I_n  L}\mu^\beta_K(y) \Big)=0
\eea
From the above Lemmas we shall derive the following Theorem. 

{\thm
\label{thm1}
On a compact Riemann surface of  genus $h \geq 1$, the following integral relations on the tensor functions $W_{I_1 \cdots I_n L}^{J_1 \cdots J_n}(x,y)$ hold, 
\bea
&&
\int_{\Sigma} W_{I_1 \cdots I_n L}^{J_1 \cdots J_n}(x,x) \, \oom^M(x) 
- \int _{\Sigma^2} \mu_L^{J_n} (y) \, W_{I_n I_{n-1} \cdots I_1}^{J_{n-1} \, \cdots \, J_1}(y,x) \, \oom^M(x)  \, \cG(x,y) 
\label{lemma4a}\\ && \quad
= 
-  \int _{\Sigma^2} \mu_\beta^{J_n} (y) \, W_{I_n I_{n-1} \cdots I_1}^{J_{n-1} \, \cdots \, J_1}(y,x) \, \oom^M(x) \,\Phi_L^\beta(x) 
\no  
\eea
The terms on the left side contain one-loop graphs, while the right side reduces to tree-level graphs only.}

\sm

To prove Theorem \ref{thm1}, we evaluate the following integral,
\bea
{ 1 \over 2 \pi i} \int _{\Sigma ^2 } \p_x \pby \cG(x,y) \, \oom^M (x) \, W_{I_1 \cdots I_n L}^{J_1 \cdots J_n}(x,y)
= 
{ 1 \over 2 \pi i} \int _{\Sigma ^2 }  \cG(x,y) \, \oom^M (x) \, \p_x \pby W_{I_1 \cdots I_n L}^{J_1 \cdots J_n}(x,y)
\qquad
\eea
in two different ways, first by the left side using Lemma \ref{Ara2} and then by the right side
using Lemma \ref{lemma3a}. The left side gives, 
\bea
-  \int _{\Sigma} W_{I_1 \cdots I_n L}^{J_1 \cdots J_n}(x,x) \, \oom ^M(x)
-  \int _{\Sigma ^2 } \mu_\beta^M (x) \, W_{I_1 \cdots I_n L}^{J_1 \cdots J_n}(x,y) \, \oom ^\beta(y)
\eea
whose second term vanishes by Lemma \ref{lemma7}.
To compute the right side, we use (\ref{lemma3a}) to evaluate the mixed derivatives of $W$, and we find,
\bea
\int_{\Sigma ^2 } \mu_L ^{J_n}(y) W^{J_{n-1} \cdots J_1}_{\,I_n \cdots I_2 I_1} (y,x) \,  \oom^M(x)  \cG(x,y)
- \int _{\Sigma ^2 } \mu_\beta^{J_n} (z) W^{J_{n-1} \cdots J_1}_{\, I_n \cdots I_2 I_1} (z,x) \oom^M(x) 
\Phi _L^\beta (x)
\eea
Equating the two gives (\ref{lemma4a}) and completes the proof of Theorem \ref{thm1}.

\sm

For weight 2, namely $n = 1$, Theorem \ref{thm1} reduces to,
\bea
&&
\int _{\Sigma} W_{I L}^{J}(x,x) \, \oom^M(x) 
- \int _{\Sigma^2} \mu_L^{J} (y) \, W_{I}(y,x) \, \oom^M(x)  \, \cG(x,y) 
\no \\ && \quad
= 
-  \int _{\Sigma^2} \mu_\beta^{J} (y) \, W_{I}(y,x) \, \oom^M(x) \,\Phi_L^\beta(x) 
\eea
Using the equations  (\ref{lemma1b}) and (\ref{otherW2}) to express $W$ in terms of the Green function and Abelian differentials, and expressing the resulting integrals in terms of the tensors $\cA$ and $\cB$, we reproduce the eight-term identity (\ref{weight2}).

\sm

At higher weight, one can similarly derive the identities (\ref{ids.12}) among modular graph tensors
by rearranging (\ref{lemma4a}) in the following way:
\begin{align}
\int_{\Sigma} W_{I_1 \cdots I_n L}^{J_1 \cdots J_n}(x,x) \, \oom^M(x) 
= \frac{i}{2}  \int _{\Sigma^2} W_{I_n I_{n-1} \cdots I_1}^{J_{n-1} \, \cdots \, J_1}(y,x) \, \oom^M(x) 
W_L(x,y) \oom^{J_n}(y)
\end{align}
The left-hand side evaluates to $-2i \cT^{J_1 J_2 \ldots J_n M}_{I_1 I_2 \ldots \, I_n L}$
as one can check by means of the representations (\ref{CDS.7}) and (\ref{ids.11}) of the $W$- and $\cT$-tensors.
The right-hand side in turn can be rewritten as $\int_{\Sigma_y} W_{I_n I_{n-1} \cdots I_1L}^{J_{n-1} \, \cdots \, J_1M}(y,y)  \oom^{J_n}(y)$ after performing the integral over $x$ through the recursive
definition of the $W$-tensors. Hence, the right-hand side is simply the relabelling
$(I_1\ldots I_NL ) \rightarrow (I_n I_{n-1}\ldots I_1 L)$ and $( J_1 \ldots J_n M) \rightarrow (J_{n-1}\ldots J_1 M J_n)$ of the left-hand side, and Theorem \ref{thm1} ultimately relates two permutations
of the $\cT$-tensor as in (\ref{ids.12}).

\subsection{The generalized triangle move}
\label{sec:4.2}

The opening line (\ref{thm0.11}) for the proof of Theorem \ref{thm0} does not rely
on any property that is specific to the Arakelov Green functions in the integrand.
The manipulations in passing to (\ref{thm0.12}) can therefore be readily adapted
to arbitrary products $F(x,z)H(y,z)$ that are $(0,1)$ forms in $x$ and $y$ as well as $(1,0)$ forms in $z$
in the place of $\cG(x,z) \cG(y,z) \overline \omega^M(x) \overline \omega^N(y) \omega_K(z) $:
\begin{align}
&\int_{\Sigma^3} \big(\partial_x \partial_{\bar z} F(x,z) \big)  H(y,z) W^{J_1 \ldots J_n}_{I_1 \ldots I_n L}(x,y)
\label{triag.1}
\\
&\ \ = \int_{\Sigma^3}  F(x,z) \big(\partial_y \partial_{\bar z} H(y,z) \big)   W^{J_n \ldots J_1}_{L I_n \ldots I_1}(y,x)
\notag
\end{align}
This follows from the same integrations parts in all of $x,y,z$ combined with
the generalized interchange lemma (\ref{lemma2a}) that led to (\ref{thm0.12}).
When visualizing all of $F,H,W$ through an edge connecting the respective
points $x,y,z$ on the surface, one can view (\ref{triag.1}) as moving the pair
of derivatives $\partial_x \partial_{\bar z} $ through the triangle graph in the following figure:
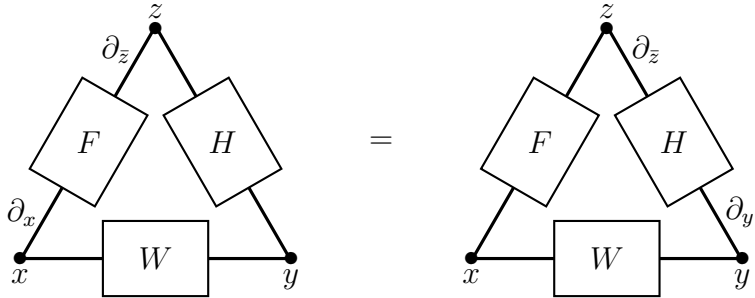
\begin{center}
\tikzpicture[scale=1]
\draw(-1.8,0)node{$\bullet$}node[below]{$x$};
\draw(1.8,0)node{$\bullet$}node[below]{$y$};
\draw(0,3.05)node{$\bullet$}node[above]{$z$};
\draw[thick](-0.7,-0.5) rectangle (0.7,0.5);
\draw(0,0)node{$W$};
\draw[very thick](-1.8,0) -- (-0.7,0);
\draw[very thick](1.8,0) -- (0.7,0);
\scope[xshift=-0.89cm, yshift=1.55cm]
\scope[rotate= 60]
\draw[thick](-0.7,-0.5) rectangle (0.7,0.5);
\draw(0,0)node{$F$};
\draw[very thick](-1.8,0) -- (-0.7,0);
\draw[very thick](1.8,0) -- (0.7,0);
\draw(1.25,0.3)node{$ \partial_{\bar z}$};
\draw(-1.25,0.3)node{$ \partial_{x}$};
\endscope
\endscope
\scope[xshift=0.89cm, yshift=1.55cm]
\scope[rotate= -60]
\draw[thick](-0.7,-0.5) rectangle (0.7,0.5);
\draw(0,0)node{$H$};
\draw[very thick](-1.8,0) -- (-0.7,0);
\draw[very thick](1.8,0) -- (0.7,0);
\endscope
\endscope
\draw(3,1.55)node{$=$};
\scope[xshift=6cm]
\draw(-1.8,0)node{$\bullet$}node[below]{$x$};
\draw(1.8,0)node{$\bullet$}node[below]{$y$};
\draw(0,3.05)node{$\bullet$}node[above]{$z$};
\draw[thick](-0.7,-0.5) rectangle (0.7,0.5);
\draw(0,0)node{$W$};
\draw[very thick](-1.8,0) -- (-0.7,0);
\draw[very thick](1.8,0) -- (0.7,0);
\scope[xshift=-0.89cm, yshift=1.55cm]
\scope[rotate= 60]
\draw[thick](-0.7,-0.5) rectangle (0.7,0.5);
\draw(0,0)node{$F$};
\draw[very thick](-1.8,0) -- (-0.7,0);
\draw[very thick](1.8,0) -- (0.7,0);
\endscope
\endscope
\scope[xshift=0.89cm, yshift=1.55cm]
\scope[rotate= -60]
\draw[thick](-0.7,-0.5) rectangle (0.7,0.5);
\draw(0,0)node{$H$};
\draw[very thick](-1.8,0) -- (-0.7,0);
\draw[very thick](1.8,0) -- (0.7,0);
\draw(1.25,0.3)node{$ \partial_{y}$};
\draw(-1.25,0.3)node{$\partial_{\bar z}$};
\endscope
\endscope
\endscope
\endtikzpicture
\end{center}
In contrast to the graphical notation of section \ref{sec:graph}, we are not keeping track of the indices here since
the main focus is on the arrangement of the derivatives.

\sm

Note that the scope of (\ref{triag.1}) goes far beyond Theorem \ref{thm0} since
$F(x,z)$ and $H(y,z)$ may be chosen to be combinations of Green functions
of arbitrary tensor rank, involving further integration points and associated with 
graphs of more general  topologies than linear chains.

\subsection{The generalized square move}
\label{sec:4.3}

Let $F(x,y,z,u)$ be a $(0,1)$ form in $x,y$ and a $(1,0)$ form in $z,u$,
then the generalized interchange lemma and its complex conjugate imply that
\begin{align}
&\int_{\Sigma^4} \big(\partial_x \partial_{\bar u} F(x,y,z,u) \big) W^{J_1\ldots J_n}_{I_1\ldots I_n L}(x,y) 
\overline{W}^{Q_1\ldots Q_\ell M}_{P_1\ldots P_\ell}(u,z)  \label{squ.1} \\
&\ \ = \int_{\Sigma^4} \big(\partial_y \partial_{\bar z} F(x,y,z,u) \big) W^{J_n\ldots J_1}_{L I_n\ldots I_1}(y,z) 
\overline{W}^{M Q_\ell \ldots Q_1}_{P_\ell \ldots P_1}(z,u) \notag
\end{align}
When the $W$-tensors are visualized through edges connecting pairs of points
$x,y$ and $z,u$ on the surface, then (\ref{squ.1}) amounts to moving the pair
of derivatives $\partial_x \partial_{\bar u} $ through a four-vertex graph as depicted in the following figure
(again dropping indices to avoid cluttering).
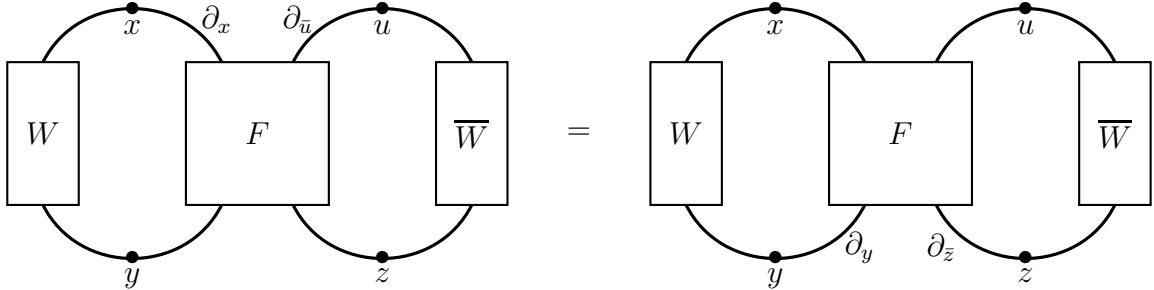
\begin{center}
\tikzpicture[scale=0.95]
\draw(-1.75,1.74)node{$\bullet$}node[below]{$x$};
\draw(1.75,1.74)node{$\bullet$}node[below]{$u$};
\draw(-1.75,-1.74)node{$\bullet$}node[below]{$y$};
\draw(1.75,-1.74)node{$\bullet$}node[below]{$z$};
\draw[very thick] (-0.5,1) .. controls (-1,2) and (-2.5,2) .. (-3,1);
\draw[very thick] (0.5,1) .. controls (1,2) and (2.5,2) .. (3,1);
\draw[very thick] (-0.5,-1) .. controls (-1,-2) and (-2.5,-2) .. (-3,-1);
\draw[very thick] (0.5,-1) .. controls (1,-2) and (2.5,-2) .. (3,-1);
\draw[thick](-3.5,1) rectangle (-2.5,-1);
\draw(-3,0)node{$W$};
\draw[thick](-1,1) rectangle (1,-1);
\draw(0,0)node{$F$};
\draw[thick](3.5,1) rectangle (2.5,-1);
\draw(3,0)node{$\overline W$};
\draw(-0.57,1.6)node{$\partial_x$};
\draw(0.57,1.6)node{$\partial_{\bar u}$};
\draw(4.5,0)node{$=$};
\scope[xshift=9cm]
\draw(-1.75,1.74)node{$\bullet$}node[below]{$x$};
\draw(1.75,1.74)node{$\bullet$}node[below]{$u$};
\draw(-1.75,-1.74)node{$\bullet$}node[below]{$y$};
\draw(1.75,-1.74)node{$\bullet$}node[below]{$z$};
\draw[very thick] (-0.5,1) .. controls (-1,2) and (-2.5,2) .. (-3,1);
\draw[very thick] (0.5,1) .. controls (1,2) and (2.5,2) .. (3,1);
\draw[very thick] (-0.5,-1) .. controls (-1,-2) and (-2.5,-2) .. (-3,-1);
\draw[very thick] (0.5,-1) .. controls (1,-2) and (2.5,-2) .. (3,-1);
\draw[thick](-3.5,1) rectangle (-2.5,-1);
\draw(-3,0)node{$W$};
\draw[thick](-1,1) rectangle (1,-1);
\draw(0,0)node{$F$};
\draw[thick](3.5,1) rectangle (2.5,-1);
\draw(3,0)node{$\overline W$};
\draw(-0.57,-1.6)node{$\partial_y$};
\draw(0.57,-1.6)node{$\partial_{\bar z}$};
\endscope
\endtikzpicture
\end{center}
With the special choice $F(x,y,z,u) = \cG(x,u) \cG(y,z) \overline \omega^A(x) \omega_B(z) \overline \omega^C(y) \omega_D(u)$ and the Laplace equation of the Green function, (\ref{squ.1}) moves
derivatives through a square graph and implies that the tensor $\cU$ defined by,
\begin{align}
&\cU^{AC \, , \, J_1\ldots J_n \, , \, Q_1\ldots Q_\ell M}_{DB \, , \, I_1\ldots I_n L \, , \, P_1 \ldots P_\ell} = 
\int_{\Sigma^3} \cG(y,z) \mu_D^A(x) \omega_B(z) \overline \omega^C(y)  W^{J_1\ldots J_n}_{I_1\ldots I_n L}(x,y) 
\overline{W}^{Q_1\ldots Q_\ell M}_{P_1\ldots P_\ell}(x,z) 
\notag\\
& \ \ \ \ - \int_{\Sigma^4} \cG(y,z) \mu_E^A(x) \mu^E_D(u)  \omega_B(z) \overline \omega^C(y) 
W^{J_1\ldots J_n}_{I_1\ldots I_n L}(x,y) 
\overline{W}^{Q_1\ldots Q_\ell M}_{P_1\ldots P_\ell}(u,z) 
 \label{squ.2}
\end{align}
obeys the following symmetry properties:
\begin{align}
\cU^{AC \, , \, J_1\ldots J_n \, , \, Q_1\ldots Q_\ell M}_{DB \, , \, I_1\ldots I_n L \, , \, P_1 \ldots P_\ell}
= \cU^{CA \, , \, J_n\ldots J_1 \, , \, M Q_\ell \ldots Q_1}_{BD \, , \, L I_n\ldots I_1 \, , \, P_\ell \ldots P_1}
 \label{squ.3}
\end{align}
The right-hand side has the pairs of indices $A \leftrightarrow C$ and $B \leftrightarrow D$ swapped
as well as all of $(I_1\ldots I_n L) , \, (J_1\ldots J_n),\, (Q_1\ldots Q_\ell M)$ and $(P_1\ldots P_\ell)$ reversed in comparison to the left-hand side. In the special case with $n=\ell=0$, the tensor (\ref{squ.2}) evaluates to
\begin{align}
\cU^{AC\, , \, M}_{DB \, , \, L} = -4 \overline \cT^{AMC}_{DBL} 
\end{align}
such that (\ref{squ.3}) simply reproduces  $\overline \cT^{AMC}_{DBL} =\overline \cT^{CMA}_{BDL}$, i.e.\
the complex conjugate of (\ref{weight3}).

\sm 

Note that $F(x,y,z,u)$ can again be chosen as combinations of Green functions of arbitrary tensor rank that may give rise to graphs of more general  topologies than linear chains. Hence, (\ref{squ.2}) should be broadly applicable to the derivation of further families of identities among modular graph tensors in future work.

\newpage

\section{Conclusion and outlook}
\label{sec:last}
\setcounter{equation}{0}

In this work, we have introduced the notion of modular graph tensors generalizing ideas of Kawazumi, and 
initiated the systematic study of identities between them. Our main result is the all-weight
family of new algebraic identities involving tree-level and one-loop graphs in (\ref{ids.11}) and (\ref{ids.12}).
Their traces over the free indices yield identities among the higher-genus modular graph functions introduced in \cite{DHoker:2017pvk}. The new identities are derived from the interchange lemma (\ref{lemma2a}) applied to suitable combinations of Arakelov Green functions.

\sm

There are several directions along which the present work may be naturally generalized. 
\begin{itemize}

\vspace{-0.1cm}
\item First, instead of integrating combinations of Arakelov Green functions against the volume forms $\mu_I^J(z)$ built out of holomorphic and anti-holomorphic one-forms, one may include differentials of the Arakelov Green function such as $dz \, \p_z \cG(z,w)$. Such objects were encountered already for genus two and low weight in \cite{DHoker:2020tcq}. 

\vspace{-0.28cm}
\item Second, a natural extension of this investigation is the application of the interchange lemma to the evaluation of the  differential with respect to moduli, and the Laplacian on moduli space,  of general classes of modular graph tensors,  and the derivation of any new  identities between such differentials and Laplacians.  This would amount to a higher-genus generalization of  modular graph forms \cite{DHoker:2016mwo} that appear in the moduli derivatives of  modular graph functions at genus one. Studies of this type have been initiated  for genus two in the context of the Laplace equation of the Zhang-Kawazumi invariant \cite{DHoker:2014oxd} and those of weight-two modular graph functions \cite{Basu:2018bde}. 

\vspace{-0.28cm}
\item Third, differential identities among modular graph tensors should lead to further  new identities upon considering their 
separating and non-separating degenerations, which may be carried out systematically  using the methods of \cite{DHoker:2017pvk, DHoker:2018mys}. One example of such a new identity was already obtained  in \cite{DHoker:2020tcq} by degenerating the genus two identity (\ref{coro.7}) to a  genus one elliptic modular graph function, and is generalized in \cite{DKS}. This genus one identity was proven directly with genus one methods in \cite{Basu:2020pey}.
\end{itemize}

\sm

Relatedly, it would be interesting to investigate the Koba-Nielsen-type integrals
in higher-genus amplitudes of the Heterotic string as tentative generating functions of more
general modular graph tensors introduced by moduli derivatives. At genus one, this kind 
of embedding of modular graph forms into Koba-Nielsen integrals has greatly advanced 
the structural understanding of modular graph functions and led to new methods for low 
energy expansions of string amplitudes \cite{Gerken:2018jrq, Gerken:2019cxz, Gerken:2020yii}.
Similar techniques should be applicable to configuration-space integrals in higher-genus 
string amplitudes and are hoped to eventually reveal suitable bases and the explicit moduli 
dependence of modular graph tensors.

\appendix 

\newpage

\section{Proof of (\ref{CDS.7})}
\label{app:A}
\setcounter{equation}{0}

In this appendix, we will prove by induction that the recursive definition 
(\ref{VWn}) of the $W$-tensors leads to the explicit formula (\ref{CDS.7}).
The latter is established for $n\leq 3$ by the examples of $W^{J_1\ldots J_n}_{I_1\ldots I_nL}(x,y)$ 
encoded in (\ref{CDS.2}) to (\ref{CDS.4}), so it remains to carry out the inductive step. Assuming
(\ref{CDS.7}) to hold at given $n$, then we will deduce the validity of the corresponding
relation at $n\rightarrow n{+}1$
from the recursion (\ref{VWn}):
\begin{align}
W^{J_1 \cdots J_n J_{n+1}} _{I_1 \cdots I_n I_{n+1} L} (x,y)  &=
 \frac{i}{2} \int_{\Sigma_z} W^{J_1 \cdots J_n} _{I_1 \cdots I_n I_{n+1}} (x,z) \overline \omega^{J_{n+1}}(z) W_L(z,y)
 \notag \\
&= \int_{\Sigma_z}  \bigg\{
V^{J_1 J_2 \dots J_n}_{\, I_1 I_2\, \ldots \, I_n}(x,z)\mu_{I_{n+1}}^{J_{n+1}}(z)
- \Phi^{J_1 J_2 \dots J_n K}_{ \, I_1 I_2\,  \ldots \, I_n I_{n+1}}(x) \mu^{J_{n+1}}_K(z)  
\no \\ 
& \ \ \ \ + \sum_{k=1}^n (-1)^k \!\!\!
\sum_{1\leq i_1<i_2<\ldots <i_k \leq n} \!\!\!
\Phi^{J_1  \ldots J_{i_1-1} \alpha_1}_{ I_1\, \ldots I_{i_1-1} I_{i_1}}(x) \, 
\cA^{J_{i_1} J_{i_1+1} \ldots J_{i_2-1} \alpha_2}_{\, \alpha_1 \, I_{i_1+1} \ldots I_{i_2-1} I_{i_2} }
\no \\ 
& \ \ \ \ \qquad  \qquad \times 
\cA^{J_{i_2} J_{i_2+1} \ldots J_{i_3-1} \alpha_3}_{\, \alpha_2 \, I_{i_2+1} \ldots I_{i_3-1} I_{i_3} }  
~ \times \cdots \times ~
\cA^{J_{i_{k-1}} J_{i_{k-1}+1} \ldots J_{i_k-1} \alpha_k}_{\, \alpha_{k-1} \, I_{i_{k-1}+1} \ldots I_{i_k-1} I_{i_k} } 
\no \\ 
&\ \ \ \ \qquad \qquad \times 
\Big (  
\Phi^{J_n  J_{n-1} \ldots  J_{i_k+1} J_{i_k} }_{ I_n \, I_{n-1} \ldots \,  I_{i_k+1}  \, \alpha_k \,} (z) \mu_{I_{n+1}}^{J_{n+1}}(z) 
- \cA^{J_{i_k} J_{i_k+1} \ldots J_n K}_{\, \alpha_k \, I_{i_k+1} \ldots \, I_n  I_{n+1}}\mu_K^{J_{n+1}}(z) \Big ) \bigg\}
\notag \\
&\ \ \times  \Big( \cG(z,y) \omega_L(y) - \Phi_L^\alpha(z) \omega_\alpha(y) \Big) 
\label{appa.1}
\end{align}
Let us separately evaluate the contributions of the first term $\sim  \cG(z,y)$ and the 
second term $\sim \Phi_L^\alpha(z) $ in the last line: Based on the recursive definitions
(\ref{VWn}) of various tensors, we have
\begin{align}
W^{J_1 \cdots J_n J_{n+1}} _{I_1 \cdots I_n  I_{n+1} L} (x,y) \, \big|_{ \cG(z,y) } &=
\underbrace{ V^{J_1 J_2 \dots J_nJ_{n+1}}_{\, I_1 I_2\, \ldots \, I_n \, I_{n+1}}(x,y) \omega_L(z) }_{(a)}
- \underbrace{ \Phi^{J_1 J_2 \dots J_n K}_{ \, I_1 I_2\,  \ldots \, I_n I_{n+1}}(x) \Phi^{J_{n+1}}_K(y) \omega_L(y)  }_{(b)}
\no \\ 
& \ \  + \sum_{k=1}^n (-1)^k \!\!\!
\sum_{1\leq i_1<i_2<\ldots <i_k \leq n} \!\!\!
\Phi^{J_1  \ldots J_{i_1-1} \alpha_1}_{ I_1\, \ldots I_{i_1-1} I_{i_1}}(x) \, 
\cA^{J_{i_1} J_{i_1+1} \ldots J_{i_2-1} \alpha_2}_{\, \alpha_1 \, I_{i_1+1} \ldots I_{i_2-1} I_{i_2} }
\no \\ 
& \ \   \qquad \times 
\cA^{J_{i_2} J_{i_2+1} \ldots J_{i_3-1} \alpha_3}_{\, \alpha_2 \, I_{i_2+1} \ldots I_{i_3-1} I_{i_3} }  
~ \times \cdots \times ~
\cA^{J_{i_{k-1}} J_{i_{k-1}+1} \ldots J_{i_k-1} \alpha_k}_{\, \alpha_{k-1} \, I_{i_{k-1}+1} \ldots I_{i_k-1} I_{i_k} } 
\label{appa.2}\\ 
&\ \  \qquad \times 
\Big (  
 \underbrace{ \Phi^{ J_{n+1} J_n    \ldots  J_{i_k+1} J_{i_k} }_{ \, I_{n+1} I_n   \ldots \,  I_{i_k+1}  \, \alpha_k \,} (y) \omega_L(y) }_{(c)}
- \underbrace{  \cA^{J_{i_k} J_{i_k+1} \ldots J_n K}_{\, \alpha_k \, I_{i_k+1} \ldots \, I_n  I_{n+1}}
\Phi^{J_{n+1}}_K(y) \omega_L(y) }_{(d)} \Big ) 
\notag
\end{align}
as well as
\begin{align}
W^{J_1 \cdots J_n J_{n+1}} _{I_1 \cdots I_n I_{n+1} L} (x,y) \, \big|_{ \Phi^\alpha_L(z) }  &=
\underbrace{ - \Phi^{J_1 J_2 \dots J_n J_{n+1} \alpha}_{\, I_1 I_2\, \ldots \, I_n I_{n+1} L}(x)
\omega_\alpha(y)}_{(e)}
+ \underbrace{ \Phi^{J_1 J_2 \dots J_n K}_{ \, I_1 I_2\,  \ldots \, I_n I_{n+1}}(x) \cA^{J_{n+1} \alpha}_{\  K \  L} \omega_\alpha(y) }_{(f)}
\no \\ 
& \ \ - \sum_{k=1}^n (-1)^k \!\!\!
\sum_{1\leq i_1<i_2<\ldots <i_k \leq n} \!\!\!
\Phi^{J_1  \ldots J_{i_1-1} \alpha_1}_{ I_1\, \ldots I_{i_1-1} I_{i_1}}(x) \, 
\cA^{J_{i_1} J_{i_1+1} \ldots J_{i_2-1} \alpha_2}_{\, \alpha_1 \, I_{i_1+1} \ldots I_{i_2-1} I_{i_2} }
\no \\ 
& \ \   \qquad \times 
\cA^{J_{i_2} J_{i_2+1} \ldots J_{i_3-1} \alpha_3}_{\, \alpha_2 \, I_{i_2+1} \ldots I_{i_3-1} I_{i_3} }  
~ \times \cdots \times ~
\cA^{J_{i_{k-1}} J_{i_{k-1}+1} \ldots J_{i_k-1} \alpha_k}_{\, \alpha_{k-1} \, I_{i_{k-1}+1} \ldots I_{i_k-1} I_{i_k} } 
\label{appa.3} \\ 
&\ \   \qquad \times 
\Big (  
\underbrace{ \cA^{ J_{i_k} J_{i_k+1} \ldots J_n J_{n+1} \alpha}_{\alpha_k  \,  I_{i_k+1} \ldots \, I_n I_{n+1} L} \omega_\alpha(y) }_{(g)}
- \underbrace{  \cA^{J_{i_k} J_{i_k+1} \ldots J_n K}_{\, \alpha_k \, I_{i_k+1} \ldots \, I_n  I_{n+1}}
\cA^{J_{n+1} \alpha}_{\ K \ L} \omega_\alpha(y) }_{(h)} \Big) 
\notag
\end{align}
In order to complete the inductive proof, we need to show that (\ref{appa.2}) and (\ref{appa.3})
add up to (\ref{CDS.7}) with $n$ shifted to $n{+}1$,
\bea
W^{J_1 \cdots J_n J_{n+1}} _{I_1 \cdots I_n I_{n+1} L} (x,y)  &=&  
\underbrace{V^{J_1 J_2 \dots J_n J_{n+1}}_{\, I_1 I_2\, \ldots \, I_n I_{n+1}}(x,y)\omega_L(y)}_{(a)}
\underbrace{- \Phi^{J_1 J_2 \dots J_n J_{n+1} K}_{ \, I_1 I_2\,  \ldots \, I_n I_{n+1} L}(x) \omega_K(y)  }_{(e)}
\no \\ &&
+ \sum_{k=1}^{n+1} (-1)^k \!\!\!
\sum_{1\leq i_1<i_2<\ldots <i_k \leq n+1} \!\!\!
\Phi^{J_1  \ldots J_{i_1-1} \alpha_1}_{ I_1\, \ldots I_{i_1-1} I_{i_1}}(x) \, 
\cA^{J_{i_1} J_{i_1+1} \ldots J_{i_2-1} \alpha_2}_{\, \alpha_1 \, I_{i_1+1} \ldots I_{i_2-1} I_{i_2} }
\no \\ && \qquad   \times 
\cA^{J_{i_2} J_{i_2+1} \ldots J_{i_3-1} \alpha_3}_{\, \alpha_2 \, I_{i_2+1} \ldots I_{i_3-1} I_{i_3} }  
~ \times \cdots \times ~
\cA^{J_{i_{k-1}} J_{i_{k-1}+1} \ldots J_{i_k-1} \alpha_k}_{\, \alpha_{k-1} \, I_{i_{k-1}+1} \ldots I_{i_k-1} I_{i_k} } 
\label{appa.4}\\ && \qquad  \times 
\Big (  
\Phi^{J_{n+1}  J_{n} \ldots  J_{i_k+1} J_{i_k} }_{ I_{n+1} \, I_{n} \ldots \,  I_{i_k+1}  \, \alpha_k \,} (y) \omega_L(y) 
- \cA^{J_{i_k} J_{i_k+1} \ldots J_{n+1} K}_{\, \alpha_k \, I_{i_k+1} \ldots \, I_{n+1}  L}\omega_K(y) \Big )
\notag
\eea
Indeed, the first line reproduces the terms $(a)$ and $(e)$ in (\ref{appa.2}) and (\ref{appa.3}),
so it remains to identify the terms $(b), (c),(d)$ and $(f),(g),(h)$ in the sum over $k$ in (\ref{appa.4}).
For this purpose, we decompose the double-sum $ \sum_{k=1}^{n+1} \sum_{1\leq i_1<i_2<\ldots <i_k \leq n+1}$ into
\begin{itemize}
\item[(i)] terms with $k=1$ and $i_k=i_1=n{+}1$
\item[(ii)] terms with $i_k = n{+}1$ and $k=2,3,\ldots,n{+}1$
\item[(iii)] terms with $i_k\neq n{+}1$ which is only possible for $k\leq n$
\end{itemize}
The first class of terms in the target expression (\ref{appa.4}) is easily seen to give
\begin{align}
W^{J_1 \cdots J_{n+1}} _{I_1 \cdots I_{n+1} L} (x,y)  \, \big|_{\text{(i)}} &= 
\underbrace{ - \Phi^{J_1 \ldots J_n \alpha}_{\, I_1\ldots \, I_n I_{n+1}}(x) \Phi^{J_{n+1}}_\alpha(y) \omega_L(y) }_{(b)}
+ \underbrace{ \Phi^{J_1 \ldots J_n \alpha}_{\, I_1\ldots \, I_n I_{n+1}}(x) \, \cA_{\ \alpha \ L}^{J_{n+1} K} \omega_K(y) }_{(f)}
\label{appa.5}
\end{align}
We next consider the second class of terms
\begin{align}
&W^{J_1 \cdots J_{n+1}} _{I_1 \cdots I_{n+1} L} (x,y)  \, \big|_{\text{(ii)}}= 
 \sum_{k=2}^{n+1} (-1)^k \!\!\!
\sum_{1\leq i_1<i_2<\ldots <i_{k-1} \leq n} \!\!\!
\Phi^{J_1  \ldots J_{i_1-1} \alpha_1}_{ I_1\, \ldots I_{i_1-1} I_{i_1}}(x) \, 
\cA^{J_{i_1} J_{i_1+1} \ldots J_{i_2-1} \alpha_2}_{\, \alpha_1 \, I_{i_1+1} \ldots I_{i_2-1} I_{i_2} }
\label{appa.6} \\
&\ \ \ \  \times 
\cA^{J_{i_2} J_{i_2+1} \ldots J_{i_3-1} \alpha_3}_{\, \alpha_2 \, I_{i_2+1} \ldots I_{i_3-1} I_{i_3} }  
~ \times \cdots \times ~
\cA^{J_{i_{k-1}} J_{i_{k-1}+1} \ldots J_{n} \alpha_k}_{\, \alpha_{k-1} \, I_{i_{k-1}+1} \ldots I_{n} I_{n+1} } 
\Big (  
\Phi^{J_{n+1} }_{\, \alpha_k } (y) \omega_L(y) 
- \cA^{ J_{n+1} K}_{\ \alpha_k \  L}\omega_K(y) \Big )
\notag
\end{align}
which can be lined up with terms $(d)$ and $(h)$ in (\ref{appa.2}) and (\ref{appa.3})
through a change of summation variable $k \rightarrow k{+}1$:
\begin{align}
&W^{J_1 \cdots J_{n+1}} _{I_1 \cdots I_{n+1} L} (x,y)  \, \big|_{\text{(ii)}}= 
- \sum_{k=1}^{n} (-1)^k \!\!\!
\sum_{1\leq i_1<i_2<\ldots <i_{k} \leq n} \!\!\!
\Phi^{J_1  \ldots J_{i_1-1} \alpha_1}_{ I_1\, \ldots I_{i_1-1} I_{i_1}}(x) \, 
\cA^{J_{i_1} J_{i_1+1} \ldots J_{i_2-1} \alpha_2}_{\, \alpha_1 \, I_{i_1+1} \ldots I_{i_2-1} I_{i_2} }
\label{appa.7} \\
&\ \ \ \  \times 
\cA^{J_{i_2} J_{i_2+1} \ldots J_{i_3-1} \alpha_3}_{\, \alpha_2 \, I_{i_2+1} \ldots I_{i_3-1} I_{i_3} }  
~ \times \cdots \times ~
\cA^{J_{i_{k}} J_{i_{k}+1} \ldots J_{n} \beta}_{\, \alpha_{k} \, I_{i_{k}+1} \ldots I_{n} I_{n+1} } 
\Big (  
\underbrace{ \Phi^{J_{n+1} }_{\, \beta } (y) \omega_L(y)  }_{(d)}
\underbrace{ - \cA^{ J_{n+1} K}_{\  \beta \  \ L}\omega_K(y) }_{(h)} \Big )
\notag
\end{align}
Finally, we recover $(c)$ and $(g)$ in (\ref{appa.2}) and (\ref{appa.3})
from the third class of terms in (\ref{appa.4}),
\begin{align}
W^{J_1 \cdots J_{n+1}} _{I_1 \cdots I_{n+1} L} (x,y)  \, \big|_{\text{(iii)}}&=
 \sum_{k=1}^{n} (-1)^k \!\!\!
\sum_{1\leq i_1<i_2<\ldots <i_k \leq n} \!\!\!
\Phi^{J_1  \ldots J_{i_1-1} \alpha_1}_{ I_1\, \ldots I_{i_1-1} I_{i_1}}(x) \, 
\cA^{J_{i_1} J_{i_1+1} \ldots J_{i_2-1} \alpha_2}_{\, \alpha_1 \, I_{i_1+1} \ldots I_{i_2-1} I_{i_2} }
\no \\ & \qquad   \times 
\cA^{J_{i_2} J_{i_2+1} \ldots J_{i_3-1} \alpha_3}_{\, \alpha_2 \, I_{i_2+1} \ldots I_{i_3-1} I_{i_3} }  
~ \times \cdots \times ~
\cA^{J_{i_{k-1}} J_{i_{k-1}+1} \ldots J_{i_k-1} \alpha_k}_{\, \alpha_{k-1} \, I_{i_{k-1}+1} \ldots I_{i_k-1} I_{i_k} } 
\label{appa.10}\\ & \qquad  \times 
\Big (  
\underbrace{ \Phi^{J_{n+1}  J_{n} \ldots  J_{i_k+1} J_{i_k} }_{ I_{n+1} \, I_{n} \ldots \,  I_{i_k+1}  \, \alpha_k \,} (y) \omega_L(y) }_{(c)}
\underbrace{ - \cA^{J_{i_k} J_{i_k+1} \ldots J_{n+1} K}_{\, \alpha_k \, I_{i_k+1} \ldots \, I_{n+1}  L}\omega_K(y)}_{(g)} \Big )  \notag
\end{align}
where the upper limit of $\sum_{k=1}^n$ follows from the fact that $i_k \neq n{+}1$ is incompatible
with $k= n{+}1$.
In summary, we have matched all the terms $(a)$ to $(h)$ in (\ref{appa.2}) and (\ref{appa.3})
with the target expression (\ref{appa.4}) of the inductive step and thereby completed the
proof of (\ref{CDS.7}) by induction.

\sm

We note that an alternative approach to this proof can be based on the
graphical representations of the recursion (\ref{CDrecursion}) of the $C$ and $D$ tensors.

\newpage

\section{Proof of (\ref{ids.11})}
\label{app:B}
\setcounter{equation}{0}

The purpose of this appendix is to derive the expression (\ref{ids.11})
for the tensor $\cT_{\,I_1\ldots I_n}^{J_1 \ldots J_n}$ in Theorem \ref{thm0}
from the all-weight formula (\ref{CDS.7}) for the $W$-tensors. The 
integral in (\ref{defttens}) will be shown to evaluate to 
\begin{align}
&\cT^{J_1 J_2 \dots J_n NM}_{\, I_1 I_2 \ldots \, I_n LK} = \underbrace{ \cB^{J_1 J_2 \dots J_nNM}_{I_1 I_2 \ldots I_n L K} }_{(A)} \,
\underbrace{ - \, \cA^{M J_1 J_2 \ldots J_n N \alpha}_{\alpha \,I_1 \, I_2 \ldots \, I_n L K} }_{(B)} \,
\underbrace{- \, \cA^{N M J_1 J_2 \ldots J_n  \alpha}_{\alpha \,K I_1 \, I_2 \ldots \, I_n L } }_{(C)} \,
\underbrace{+ \, \cA^{M J_1 J_2 \ldots J_n \beta}_{\alpha \,I_1 \, I_2 \ldots \, I_n L } \cA^{N \alpha}_{\beta K} }_{(D)} \notag \\
&\ \ \ \ + \sum_{k=1}^n (-1)^k
\sum_{1\leq i_1<i_2<\ldots <i_k \leq n}
\cA^{J_{i_1} J_{i_1+1} \ldots J_{i_2-1} \alpha_2}_{\, \alpha_1 \, I_{i_1+1} \ldots I_{i_2-1} I_{i_2} }
\cA^{J_{i_2} J_{i_2+1} \ldots J_{i_3-1} \alpha_3}_{\, \alpha_2 \, I_{i_2+1} \ldots I_{i_3-1} I_{i_3} } 
 \times \ldots \times
\cA^{J_{i_{k-1}} J_{i_{k-1}+1} \ldots J_{i_k-1} \alpha_k}_{\, \alpha_{k-1} \, I_{i_{k-1}+1} \ldots I_{i_k-1} I_{i_k} }  \notag \\
& \ \ \ \ \ \ \ \ \times 
\Big(
\underbrace{\cA^{J_{i_k} J_{i_k+1} \ldots J_n NM J_1  \ldots J_{i_1-1} \alpha_1}_{\, \alpha_k \, I_{i_k+1} \ldots \, I_n LK \, I_1\, \ldots I_{i_1-1} I_{i_1} } }_{(E)} \,
\underbrace{ - \, \cA^{J_{i_k}  \ldots J_n N \beta}_{\, \alpha_k   \ldots \, I_n LK}  \cA^{M J_1  \ldots   \alpha_1}_{\, \beta \, I_1\, \ldots   I_{i_1} } }_{(F)} \label{appb.1} \\
 &\hspace{2cm}
\underbrace{ - \,  \cA^{J_{i_k}  \ldots J_n  \beta}_{\, \alpha_k   \ldots \, I_n L}  \cA^{N M J_1  \ldots   \alpha_1}_{\, \beta \, K I_1\, \ldots   I_{i_1} } }_{(G)}
 + \underbrace{ \cA^{J_{i_k}  \ldots J_n  \beta}_{\, \alpha_k   \ldots \, I_n L} \cA^{N \gamma}_{\beta K}  \cA^{M J_1  \ldots   \alpha_1}_{\, \gamma \; I_1\, \ldots   I_{i_1} }  }_{(H)} \Big)
\notag 
\end{align}
which is a rewriting of the target expression (\ref{ids.11}) at $n \rightarrow n{+}2$, where $NM$ and $LK$ take the
role of $J_{n+1}J_{n+2}$ and $I_{n+1}I_{n+2}$, respectively. More precisely, we have regrouped
the terms in the $(n{+}2)$-point version of (\ref{ids.11}) according to whether $\begin{smallmatrix} N \\ L \end{smallmatrix}$ or $\begin{smallmatrix} M \\ K \end{smallmatrix}$ have been replaced by contractions
$\cA^{\ldots \alpha}_{\ldots L} \cA_{\alpha \ldots }^{N \ldots}$ or
$\cA^{\ldots \alpha}_{\ldots K} \cA_{\alpha \ldots }^{M \ldots}$.
The second to fourth term in the first line of (\ref{appb.1}) appear since each term in the sum over $k$
has at least one $\begin{smallmatrix} J_i \\ I_i \end{smallmatrix}$ replaced by
$\cA^{\ldots \alpha}_{\ldots I_i} \cA_{\alpha \ldots }^{J_i \ldots}$, i.e.\ they account for
the possibility to only replace one or both of $\begin{smallmatrix} N \\ L \end{smallmatrix}$, $ \! \begin{smallmatrix} M \\ K \end{smallmatrix}$ by contractions and none of~$\begin{smallmatrix} J_i \\ I_i \end{smallmatrix}$.

In order to recover (\ref{appb.1}) from the integral in (\ref{defttens}), we employ the
representation (\ref{CDS.7}) of the $W$-tensor in the integrand to find
\begin{align}
&\cT^{J_1 J_2 \dots J_n NM}_{\, I_1 I_2 \ldots \, I_n LK}= \frac{i}{2} \int _{\Sigma^2_{x,y}} \Big ( \cG(x,y) \mu_K^M(x) 
- \mu _\a ^M(x) \Phi _K^\a(y) \Big ) W_{I_1 \cdots I_n L} ^{J_1 \cdots J_n} (x,y)  \oom^N(y)\label{appb.2} \\
&\ \ = \int _{\Sigma^2_{x,y}}  \Big ( \cG(x,y) \mu_K^M(x) 
- \mu _\a ^M(x) \Phi _K^\a(y) \Big )
\bigg\{
V^{J_1 J_2 \dots J_n}_{\, I_1 I_2\, \ldots \, I_n}(x,y)\mu^N_L(y)
- \Phi^{J_1 J_2 \dots J_n \beta}_{ \, I_1 I_2\,  \ldots \, I_n L}(x) \mu^N_\beta(y)  
\no \\ &
\ \ \ \ + \sum_{k=1}^n (-1)^k \!\!\!
\sum_{1\leq i_1<i_2<\ldots <i_k \leq n} \!\!\!
\Phi^{J_1  \ldots J_{i_1-1} \alpha_1}_{ I_1\, \ldots I_{i_1-1} I_{i_1}}(x) \, 
\cA^{J_{i_1} J_{i_1+1} \ldots J_{i_2-1} \alpha_2}_{\, \alpha_1 \, I_{i_1+1} \ldots I_{i_2-1} I_{i_2} }
 \cA^{J_{i_2} J_{i_2+1} \ldots J_{i_3-1} \alpha_3}_{\, \alpha_2 \, I_{i_2+1} \ldots I_{i_3-1} I_{i_3} }  
~ \ldots \times \cdots
\no \\ & \qquad \qquad  \times ~
\cA^{J_{i_{k-1}} J_{i_{k-1}+1} \ldots J_{i_k-1} \alpha_k}_{\, \alpha_{k-1} \, I_{i_{k-1}+1} \ldots I_{i_k-1} I_{i_k} } 
\Big (  
\Phi^{J_n  J_{n-1} \ldots  J_{i_k+1} J_{i_k} }_{ I_n \, I_{n-1} \ldots \,  I_{i_k+1}  \, \alpha_k \,} (y) \mu^N_L(y) 
- \cA^{J_{i_k} J_{i_k+1} \ldots J_n \beta}_{\, \alpha_k \, I_{i_k+1} \ldots \, I_n  L}\mu^N_\beta(y) \Big )
\bigg\} \notag  
\end{align}
The integrals on the right-hand side can all be performed by repeatedly using the recursive definitions
of the tensors $\Phi,\cA$ and $\cB$: The contributions from the Green function accompanied by 
$\mu_K^M(x)$ in the second line are
\begin{align}
&\! \! (\ref{appb.2})\, \big|_{\cG(x,y)}  =  \underbrace{ \cB^{J_1 \ldots J_n NM }_{\, I_1 \ldots \, I_n LK} }_{(A)}
\, \underbrace{ -\,  \cA^{NM J_1 \ldots J_n \beta}_{\beta K \, I_1 \ldots \, I_n L}}_{(C)} +  \sum_{k=1}^n (-1)^k \!\!\! \! \! \! \!\!
\sum_{1\leq i_1<i_2<\ldots <i_k \leq n} \!\! \! \! \! \! \! \!
\cA^{J_{i_1} J_{i_1+1} \ldots J_{i_2-1} \alpha_2}_{\, \alpha_1 \, I_{i_1+1} \ldots I_{i_2-1} I_{i_2} }
 \cA^{J_{i_2} J_{i_2+1} \ldots J_{i_3-1} \alpha_3}_{\, \alpha_2 \, I_{i_2+1} \ldots I_{i_3-1} I_{i_3} } \notag \\
 &\ \  \ldots  \cA^{J_{i_{k-1}} J_{i_{k-1}+1} \ldots J_{i_k-1} \alpha_k}_{\, \alpha_{k-1} \, I_{i_{k-1}+1} \ldots I_{i_k-1} I_{i_k} } 
 \Big( 
\underbrace{ \cA^{J_{i_k} J_{i_{k}+1} \ldots J_n NM J_1 \ldots J_{i_1-1}\alpha_1}_{\alpha_k \, I_{i_k+1}\, \ldots I_n LK \, I_1 \,\ldots \, I_{i_1-1} \, I_{i_1}} }_{(E)} \,
 \underbrace{ -\, \cA^{J_{i_k} J_{i_{k}+1}  \ldots J_n \beta}_{ \alpha_k \, I_{i_k+1} \, \ldots I_n L} \cA^{ N M J_1 \ldots J_{i_1-1} \alpha_1}_{\beta K \, I_1\ldots \, I_{i_1-1} \, I_{i_1}} }_{(G)}
 \Big)
\label{appb.3}
\end{align}
whereas the contributions from $\Phi^\alpha_K(y)$ in the second line are
\begin{align}
&(\ref{appb.2})\, \big|_{\Phi^\alpha_K(y)}  = \underbrace{-  \cA^{M J_1 \ldots J_n N \alpha }_{ \alpha \, I_1 \ldots \, I_n LK}}_{(B)}
+ \underbrace{ \cA^{N \alpha}_{\beta K} \cA^{M J_1 \ldots J_n \beta}_{\alpha  \, I_1\, \ldots \, I_n L} }_{(D)}+  \sum_{k=1}^n (-1)^k \!\!\! \! \! \! \!\!
\sum_{1\leq i_1<i_2<\ldots <i_k \leq n} \!\! \! \! \! \! \! \!
\cA^{J_{i_1} J_{i_1+1} \ldots J_{i_2-1} \alpha_2}_{\, \alpha_1 \, I_{i_1+1} \ldots I_{i_2-1} I_{i_2} }
 \notag \\
 &\ \ \ \  \ldots  \cA^{J_{i_{k-1}} J_{i_{k-1}+1} \ldots J_{i_k-1} \alpha_k}_{\, \alpha_{k-1} \, I_{i_{k-1}+1} \ldots I_{i_k-1} I_{i_k} } 
 \Big( 
\underbrace{ {-}\cA^{J_{i_k} \ldots J_n N \beta}_{\alpha_k \ldots \, I_n LK} \cA^{M J_1 \ldots \alpha_1}_{\beta \, I_1\, \ldots \, I_{i_1}} }_{(F)}
+ \underbrace{ \cA^{J_{i_k} \ldots J_n \gamma }_{\alpha_k \ldots \, I_n L} \cA^{N \beta}_{\gamma K} \cA^{M J_1 \ldots \alpha_1}_{\beta \, I_1\, \ldots \, I_{i_1}} }_{(H)}
 \Big) \label{appb.4}
\end{align}
All the desired eight terms $(A)$ to $(H)$ in (\ref{appb.1}) are reproduced by (\ref{appb.3}) and (\ref{appb.4})
which completes our proof of (\ref{ids.11}).

\sm

We note that an alternative approach to this proof can be based on the
graphical representations of the tensors ${\cal T}$, $C$ and $D$ given in
sections \ref{sec:graph} and \ref{sec:mainthm}.
 
\newpage

\providecommand{\href}[2]{#2}\begingroup\raggedright\endgroup

\end{document}